%% file: trias.tex
\documentstyle[epsf]{elsart}

\begin{document}
\begin{frontmatter}
\title{Discrete Breathers in Josephson Ladders} 

\author[MIT]{E. Tr\'{\i}as\thanksref{mail1}},
\author[ZGZ1,ZGZ2]{J. J. Mazo\thanksref{mail2}},
\author[MIT,TWE]{A. Brinkman} and
\author[MIT]{T. P. Orlando}

\address[MIT]{Department of Electrical Engineering and Computer
  Science.  Massachusetts Institute of Technology, Cambridge,
  Massachusetts 02139}
\address[ZGZ1]{Departamento de F\'{\i}sica de la Materia Condensada,
  Universidad de Zaragoza, E-50009 Zaragoza, Spain}
\address[ZGZ2]{Departamento de Teor\'{\i}a y Simulaci\'on de Sistemas
  Complejos, Instituto de Ciencia de Materiales de Arag\'on,
  CSIC-Universidad de Zaragoza, E-50009 Zaragoza, Spain}
\address[TWE]{Department of Applied Physics, University of Twente,
  P.O. Box 217, 7500AE Enschede, The Netherlands}

\thanks[mail1]{etrias@mit.edu} 
\thanks[mail2]{juanjo@posta.unizar.es}

\begin{abstract}
  We present a study of nonlinear localized excitations called discrete
  breathers in a superconducting array.  These localized solutions were
  recently observed in Josephson-junction ladder arrays by two different
  experimental groups~\cite{trias00,binder00a,binder00b}. We review the
  experiments made by Tr\'{\i}as et al.~\cite{trias00} We report the detection
  of different single-site and multi-site breather states and study the
  dynamics when changing the array bias current. By changing the temperature
  we can control the value of the damping (the Stewart-McCumber parameter) in
  the array, thus allowing an experimental study at different array
  parameters. We propose a simple DC circuit model to understand most of the
  features of the detected states. We have also compared this model and the
  experiments with simulations of the dynamics of the array.  We show that the
  study of the resonances in the ladder and the use of harmonic balance
  techniques allow for understanding of most of the numerical results. We have
  computed existence diagrams of breather solutions in our arrays, found
  resonant localized solutions and described the localized states in terms of
  vortex and antivortex motion.
\end{abstract}


\end{frontmatter}


\def\figsize{3.5in}

\section{Introduction}
\label{sec:int}

Linear models of crystals have been instrumental in developing a
physical understanding of the solid state.  Thermodynamic properties
such as specific heat, transport properties such as electron
relaxation times or superconductivity, and even interactions with radiation
can be understood by modeling a crystal as a lattice of atoms with
fixed harmonic coupling.  This leads to the conventional
phonon-like analysis with a basis of plane waves as normal modes.  However,
certain properties of solids, such as thermal expansion, cannot be
understood in this linear model.  For example, the elastic constants of the
atomic interactions may depend on the temperature or the volume and so
make the interaction non-linear.  This is termed an anharmonic effect
and the usual approach is to use a more generalized Taylor expansion
for the lattice coupling that includes more than just the harmonic
term.

Until quite recently anharmonic effects were only studied as
perturbations to the fully solvable harmonic model. Then it was
discovered~\cite{sievers88} and proved~\cite{mackay94} that in
classical Hamiltonian systems non-linearity may lead to localized vibrations in
the lattice that cannot be analyzed using the standard plane wave
approach.  These intrinsic localized modes are time periodic and
spatially localized solutions and have been termed discrete breathers
(DB).  Their amplitudes oscillate around a few sites in the lattice
and they do not depend on impurities for their localization.  The
study of DB has been extended to the dynamics of coupled rotor
lattices~\cite{takeno96} where the terms ``rotating localized modes''
or ``rotobreathers'' were introduced. DB have also been proven to
exist in the dynamics of dissipative systems~\cite{mackay98}. In this
case chaotic localized solutions have been
discovered~\cite{martinez99}. DB have been found to play an important
role in conditions far from equilibrium~\cite{tsironis96} and recently
have been studied in disordered
systems~\cite{archilla99,kopidakis00}. Excellent reviews on the topic
are~\cite{aubry97} and~\cite{flach98}.

Because DB are generic modes in many non-linear lattices, they are the
object of great theoretical and numerical attention in many diverse
fields like condensed matter
physics~\cite{sievers95,floria96,mcgurn99,lai99}, mechanical
engineering~\cite{vakakis96,vakakis99,levine93} and
biophysics~\cite{peyrard98}.  Only recently have the first experiments
been performed which detect intrinsic localized modes in
quasi-one-dimensional charge-density-wave compounds~\cite{swanson99},
antiferromagnetic anharmonic crystals~\cite{schwarz99} and
superconducting arrays~\cite{trias00,binder00a,binder00b}.

The existence of DB in a Josephson-junction (JJ) array was first proposed by
Flor\'{\i}a et al.~\cite{floria96} in the study of the dynamics of an
AC-biased anisotropic ladder array. Both, oscillating and rotating localized
modes were simulated and studied in this system~\cite{floria98,martinez98}.
Later, rotobreathers were also numerically studied in the dynamics of
inductively coupled junctions in DC-biased arrays~\cite{flach99,mazo99}. 
Arrays were then designed, fabricated and measured and DB were 
found~\cite{trias00,binder00a,binder00b}.

The Josephson junctions studied in this article are made from a
superconducting-insulator-superconductor (SIS)
tunneling structure that
because of the Josephson effect behave as solid-state non-linear
oscillators. In the framework of the resistively and capacitively
shunted junction (RCSJ) model a single JJ is modeled by a parallel
combination of an ideal junction, a capacitor $C$, and a resistance
$R$. The current of the SIS tunnel junction is then 
\begin{equation}
I=C \frac{dV}{dt}+\frac{V}{R}+I_J.
\end{equation}
The ideal junction has a constitutive relation of $I_J=I_{c}\sin \varphi$
where $\varphi$ is the gauge-invariant phase difference of the junction and
the voltage $V$ across a junction is $V=(\Phi_0 / 2 \pi) d\varphi / dt$. Thus,
the current of SIS junction, $i$, in reduced units is given by
\begin{equation}
i = \ddot{\varphi}+\Gamma \dot{\varphi}+\sin{\varphi}={\cal N}(\varphi).
\label{eq:pendulum}
\end{equation}
This current is normalized by the junction's intrinsic critical current
$I_c$ and $\Gamma$, the damping, is usually referred to as the
Stewart-McCumber parameter $\beta_c=\Gamma^{-2}=2\pi I_c C R^2 / \Phi_0$
($\Phi_0$ is the flux quantum). Time is normalized by $\tau=\sqrt{ \Phi_0 C /
  2\pi I_c}$.  The RCSJ model equation is isomorphic to the equation
of a driven pendulum.  The mass is normalized to one and the viscous
damping is $\Gamma$.

\begin{figure}[bt]
\epsfxsize=2.5in
\centering{\mbox{\epsfbox{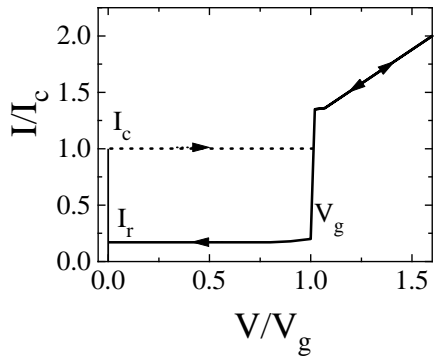}}}
%\vspace{0.15in}
\caption[Sketch]{Sketch of the IV curve of an underdamped Josephson junction.}
\label{fig:sketch}
\end{figure}

Fig.~\ref{fig:sketch} sketches a typical current-voltage (IV) curve of
a single junction. The junction is biased by a DC current and the
average voltage is measured. The IV curve of a JJ presents two
different physical states. The first is a zero-voltage state, the
so-called superconducting or quiet state; a current flows through the
junction but no voltage difference appears. This state exists for
values of the current smaller than the junction critical current
$I_c$.  At this value of the current the junction switches to the
superconducting gap voltage, $V_g$.  The gap voltage results from the
breaking of Cooper-pairs and causes the junction's resistance, and
thereby the damping, to change in a complicated nonlinear way. If we
increase the applied current further, the junction reaches its normal
state and it behaves as a resistor, $R_n$. These resistive states are
also called rotating or whirling modes. As the current decreases the
junction returns to the gap voltage and then to its zero-voltage state
at the retrapping current, $I_r$.  Thus, the resistive state occurs
for values of the current above the junction retrapping current and
coexists with the superconducting state for currents between the
critical and the retrapping values. The amplitude of this hysteretic
loop is governed by the value of the damping $\Gamma$. The behavior of
the curve (Fig.~\ref{fig:sketch}) close to $V_g$ can be modeled using
the RCSJ model with an appropriate nonlinear voltage dependence for
the resistance (see section~\ref{sub:nlr}) or by using other more
sophisticated models~\cite{likharev_book}.

We can design JJ arrays of different geometries and
parameters.  Networks of junctions are
valuable model systems for the study of coupled non-linear
oscillators.  For instance, solid-state physical realizations of the
Frenkel-Kontorova model for dislocations~\cite{floria96fk} and the
two-dimensional X-Y model for phase transitions in condensed
matter~\cite{XY} are two prominent examples.  There has also been
extensive studies on the soliton dynamics in one-dimensional Josephson
arrays~\cite{watanabe95}.  The experimental advantage of Josephson
networks is that there is good control over the parameters because
they are fabricated microelectronic solid-state circuits.
Moreover, these networks can be designed for a wide range of
oscillator parameters from the extremely underdamped to overdamped
limits.

Here we present an in-depth study of the experiments reported
in~\cite{trias00}.  We use more complete models to understand the
dynamics of the array. In particular we will not assume uniform
current bias, the effects of temperature will be discussed, a
nonlinear junction resistance is added, and a full-inductance 
matrix will be used. In the
next section we introduce the governing equations
and explain intrinsic localization in the ladder.
In section~\ref{sec:exp} we will report on the experimental study of DB
in our superconducting arrays.  We develop in section~\ref{sec:cir} a
simple circuit model which allows for the understanding of most of the
experimental findings.  Numerical
simulations will be shown section~\ref{sec:brsims}.
In section~\ref{sec:ta} we perform a linear analysis that 
yields resonance frequencies and decay lengths of
excitations.  Section~\ref{sec:nas}
is devoted to a numerical study of
single--site DB solutions in the ladder. There we compute regions of
existence of DB at different array parameters, and we will
analyze typical simulated IV curves for the cases of type-B and
type-A DB.  We also study the vortex patterns associated to both
solutions.  The major results are summarized in the final section.

\section{Josephson Ladders}

\subsection{Ladder equations}
\label{sec:ce}

\begin{figure}[bt]
\epsfxsize=\figsize
\centering{\mbox{\epsfbox{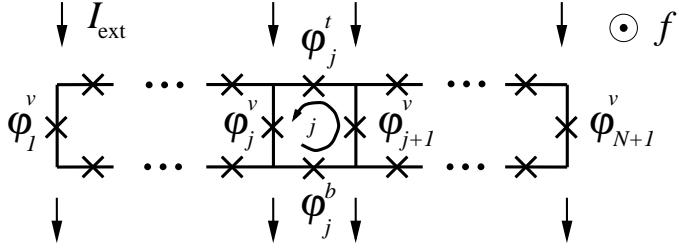}}}
%\vspace{0.15in}
\caption[Schematic of anisotropic ladder]{Anisotropic ladder array
with uniform current injection.  Vertical junctions (with superscript
$v$) have critical current $I_{cv}$ and horizontal junctions (with
superscript $t$ and $b$) have a critical current $I_{ch}$.}
\label{fig:brdiagram}
\end{figure}

Fig.~\ref{fig:brdiagram} shows the circuit diagram for the Josephson
array in ladder geometry and with uniform current injection.  The
junctions are marked by an ``$\times$''. Horizontal junctions have
critical currents of $I_{ch}$ while vertical junctions have a critical
current of $I_{cv}$. Anisotropic arrays are fabricated by varying the
area of the junctions. In our junctions, the critical current and
capacitance are proportional to this area.  Due to the constant
$I_cR_n$ product, the normal state resistance is inversely
proportional to the junction area.  The anisotropy parameter $h$ can
then be defined as $h=I_{ch}/I_{cv}=C_h/C_v=R_v/R_h$.

A ladder is a useful geometry to study DB because vertical
junctions can play the role of pendula while the
horizontal junction can act as controllable non-linear coupling.  
Thereby, a ladder can be roughly thought of as a one-dimensional
chain of pendula that are coupled by non-linear springs.  
If the phases of the vertical junctions are interpreted
as as particle coordinates, then the ladder is in essence a
one-dimensional chain of particles with non-linear
on-site potential as well as non-linear interactions since the
horizontal junctions are nonlinear Josephson elements.  
It is known that lattices with non-linear on-site potential or non-linear
interactions are needed to support DB excitations.

The junctions in the array are coupled by means of current conservation and
fluxoid quantization.
Kirchhoff's current conservation law (KCL) at the top node yields
\begin{equation}
I^t_{j-1}+I_{\rm ext}-I^t_j-I^v_j=0.
\label{eq:ladKCLbr}
\end{equation}
A consequence of the open boundary conditions is that the current on the top
horizontal junctions must be equal but opposite to the current in the bottom
horizontal junctions. Thus, $I^m_j=-I^t_j=I^b_j$.  We will normalize all the
currents by $I_{cv}$.  Also, we will
refer to horizontal junctions by the superscript $h$ (which is not to be
confused with the anisotropy) when we are dealing with either the top
or bottom horizontal junction.

Fluxoid quantization causes the circulation of the gauge invariant
phase differences around a loop to be equal to the flux of the total
(external plus induced) magnetic field through the loop.  When we
impose this condition on one of our cells and assume only external and
self-induced fields, that is, we neglect mutual inductances between
different cells, we find
\begin{equation}
\varphi^v_{j}-\varphi^v_{j+1}-\varphi^t_j+\varphi^b_j+2\pi f+{1\over
\lambda}i^m_j=0.
\label{eq:fqbrlad}
\end{equation}
Here, $f=\Phi_{\rm ext}/\Phi_0$ is the normalized applied flux
per unit cell and $\lambda=\Phi_0/2 \pi I_{cv} L_s$ where
$L_s$ is the self-inductance of the loop and
$i^m_j$ is the normalized mesh current so that $i^m_j/\lambda$ is 
the normalized self-induced flux of a cell.

The vorticity, $n_j$ is defined through the expression
\begin{equation}
[\varphi^v_j]-[\varphi^v_{j+1}]-[\varphi^t_j]+[\varphi^b_j]
=2\pi(n_j-f-f^{\rm ind}_j)
\label{eq:vorticity}
\end{equation}
where $[\varphi]$ represents the phases modulus $2\pi$, and $f^{\rm
ind}_j=i^m_j/2\pi\lambda$. This expression is equivalent
to Eq. (\ref{eq:fqbrlad}) and thus also referred to as fluxoid quantization.

We let the functional ${\cal N}(\varphi)=\ddot{\varphi}+
\Gamma\dot{\varphi} +\sin\varphi$ represents the current through a
junction in the RCSJ model.  The resulting set of nonlinear coupled
equations can be written as
\begin{eqnarray}
{\cal N} (\varphi^t_j) & = &
{\lambda \over h} \{\varphi^v_{j}-\varphi^v_{j+1}-\varphi^t_j+\varphi^b_j
+ 2 \pi f\}  \nonumber \\
{\cal N} (\varphi^v_j) & = &
\lambda \{\varphi^v_{j+1}-2\varphi^v_j+\varphi^v_{j-1}+\varphi^t_j
-\varphi^t_{j-1}-\varphi^b_j+\varphi^b_{j-1} \} + i_{\rm ext} 
\label{eq:ladder0} \nonumber \\
{\cal N} (\varphi^b_j) & = &
-{\lambda \over h} \{\varphi^v_{j}-\varphi^v_{j+1}-\varphi^t_j+\varphi^b_j
+ 2 \pi f\}
\label{eq:ladder}
\end{eqnarray}
We can identify in these equations a discrete Laplacian term $\nabla^2
\varphi^v_j=\varphi^v_{j+1}-2\varphi^v_j+\varphi^v_{j-1}$, which accounts
for the interaction between vertical junctions, and discrete first
order derivatives $\delta_x
\varphi^v_j=\varphi^v_{j+1}-\varphi^v_j$ and $\delta_x
\varphi^h_{j-1}=\varphi^h_j-\varphi^h_{j-1}$, which account for the
interaction terms between vertical and horizontal junctions respectively.

The external current is normalized as $i_{\rm ext}=I_{\rm ext}/I_{cv}$. The
damping is $\Gamma=\sqrt{\Phi_0/2\pi I_{cv} R_{v}^2C_{v}}$. We note that
because the anisotropy in our arrays is caused by varying the junction area,
$\Gamma$ is the same for every junction in the array.

In Eq. (\ref{eq:ladder}), $j=1$ to $N$ and at the open boundaries,
\begin{eqnarray}
\varphi^t_0 & = &\varphi^t_{N}=0 \nonumber \\
\varphi^b_0 & = &\varphi^b_{N}=0 \nonumber \\
\varphi^v_{N+1}& = & \varphi^v_{N} + 2\pi f \nonumber \\
\varphi^v_{0} &= &\varphi^v_{1} - 2\pi f
\end{eqnarray}
where the phases at $j=0$ and $j=N+1$ are for mathematical convenience
and do not represent real junctions.

The physical currents through the junctions are $I^t_j=I_{ch} {\cal
  N}(\varphi^t_j)= -I^b_j=-I_{ch} {\cal N}(\varphi^b_j)$ and $I^v_j=I_{cv}
{\cal N}(\varphi^v_j)$.

We can use these governing  equations to compare the ladder with a JJ
parallel array, a system with broad interest.
In a parallel array the horizontal junctions of the ladder are
replaced by superconducting wires which have a linear current-phase
relation.  In the ladder, the dynamics of the horizontal junctions can
also be described by a linearized constitutive relation under some
restricted circumstances.  For instance, in the static case, when the
horizontal junctions have no time dependence then $i^h=h\sin
\varphi^h\approx h \varphi^h$. From the normalized Eq.
(\ref{eq:ladKCLbr}),
\begin{equation}
i^v_j-i_{\rm ext}=i^t_{j-1}-i^t_j
\end{equation}
The right hand side is simply $h(\varphi^t_{j-1}-\varphi^t_{j})$
and we can find a similar relation for the bottom horizontal
junctions.  We can substitute these linearized
relations for $\varphi^t$ and
$\varphi^b$ in Eq. (\ref{eq:ladder0}) to get,
\begin{equation}
{\cal N}(\varphi^v_j)={h \lambda \over h + 2 \lambda}\nabla^2\varphi^v_j+i_{\rm ext}
\end{equation}
This is the discrete sine-Gordon equation with a renormalized
discreteness parameter of $\lambda_{\rm eff}= h \lambda / (h + 2
\lambda )$ and is equivalent to the governing equations of a
Josephson-junction parallel array~\cite{watanabe95}.

The difference between the parallel array and the ladder is obviously
the existence of horizontal junctions in the case of the ladder.  The
use of $\lambda_{\rm eff}$ to map the ladder to the discrete
sine-Gordon model is correct only for the study of a reduced set of
possible states of the array: Those for which only the convex part of
the inter-phase interaction is relevant. If we can neglect the
dynamics of the horizontal junctions (for instance when studying
static properties) then the above is a very good approximation as has
been rigorously stated in~\cite{mazo95}. {\em The rotating DB states
studied in this paper are a good example of a set of solutions of the
ladder which do not appear in the dynamics of the discrete sine-Gordon
model.}

\subsection{Localization in the Josephson ladder}

In Fig.~\ref{fig:sketch} we showed that the IV curve of an underdamped JJ has
an hysteresis loop for current values between the critical and the retrapping
currents. In this current range the zero-voltage ($V=0$) and rotating ($V =
V_g$) attractors coexist, and it is this hysteresis loop that allows for the
existence of breathers in the ladder with DC bias current.  In the full
ladder, the phase space is much more complex.  However, when the vertical
junctions are weakly coupled then it is possible that each of the junction
attractors is essentially independent.  A breather is then a localized state
where one vertical junction is rotating while the others are in the zero
voltage state.  Under general considerations of $N$ nonlinear oscillators,
it is not obvious that the phase space of the coupled system supports
attractors of localized solutions.  Indeed, in linear systems even when the
oscillators are weakly coupled, the phase space does not support stable
localized solutions.

The coupling of vertical junctions occurs through horizontal junctions,
geometrical inductances, and fluxoid quantization. But the most important
contribution is from horizontal junctions and their influence is measured by
the parameter $h$.  If the coupling is too strong, i.e. $h$ is too large, then
localized solution cannot exist. We numerically found that small $\lambda$ and
$h=0.25$ are adequate values of the coupling parameters for studying
DB~\cite{mazo99}.

\begin{figure}[tb]
\epsfxsize=\figsize
\centering{\mbox{\epsfbox{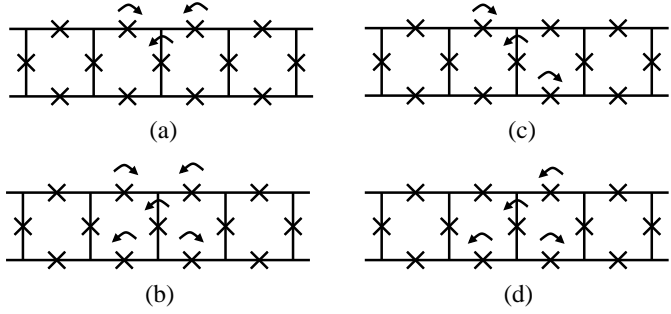}}}
\caption[Schematic of possible single-site breathers]{Different
patterns for single-site DB solutions in the ladder. Each array is
vertically biased by a constant DC current as shown in
Fig.~\ref{fig:brdiagram}.}
\label{fig:1site}
\end{figure}

Fig.~\ref{fig:1site} shows some of the possible configurations for
single-site breathers in our ladders.  The arrows indicate voltage
polarity.  Junctions without arrows are in the zero-voltage state.
The solutions in Fig.~\ref{fig:1site} are single-site breathers
because only one vertical junction is rotating.  We also see that due
to Kirchhoff's voltage law (KVL), there must be at least one other
junction rotating in each of the neighboring cells.

The actual number and pattern of rotating horizontal junctions will
determine the type of breather.  We call breathers that have two
rotating horizontal junctions, like Fig.~\ref{fig:1site}(a) and
Fig.~\ref{fig:1site}(c), type A breathers.  We see that in this case
the voltages of all the rotating junctions are the same and so the
breather solution is highly constrained.  We will use the term
asymmetric to refer to this set of type $A$ solutions.

In the case of the type $B$ solution, the four nearest horizontal junctions
rotate as depicted in Fig.~\ref{fig:1site}(b).  We will use the term symmetric
to refer to type $B$ solution. However, it is important to note that symmetric
refers only to the fact that top and bottom horizontal junction in the same
cell rotate. It does not always correspond to true up-down symmetric
solutions.  An up-down symmetric solution is a solution for which
$\varphi^t_j(t)=-\varphi^b_j(t)\,{\rm mod}\,2\pi$. This is a solution of the
system as can be inferred from Eqs. (\ref{eq:ladder}), where it is clear that
${\cal N}(\varphi^t_j)=-{\cal N}(\varphi^b_j)$. However, for certain values of
the parameters, the underdamped character of the junctions allows for
different solutions, such as the type A breathers, that do not obey the
up-down symmetry condition.  Also, for most type B solutions the magnitude of
the voltage of a horizontal rotating junction is one half the voltage of the
rotating vertical junction.  Again, not all type B solutions obey this condition.

Single-site breather solutions that have 3 horizontal junctions rotating such
as those shown in Fig.~\ref{fig:1site}(d) are called hybrids of type A and B.
The other possible single-site breathers that are not shown can also be
classified as either A, B or hybrid.

\begin{figure}[tb]
\epsfxsize=4.2in
\centering{\mbox{\epsfbox{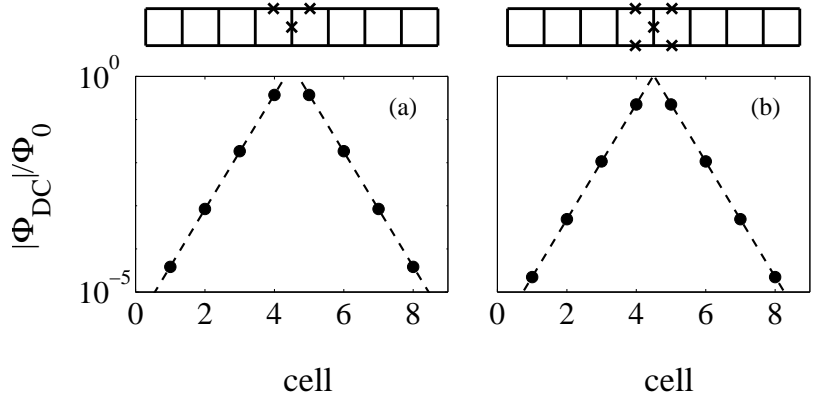}}}
\caption[Simulation of 9x1 array]{Simulation of 9x1 array with 
  $\lambda=0.05, \Gamma=0.1$ and $h=0.25$. We have plotted the absolute value of
  the DC flux per unit cell at $I=0.7$.  The flux decays exponentially with a
  decay length of 0.32 for both solutions.}
\label{fig:flux}
\end{figure}

In the rest of this paper we will designate librating
junctions as shorts and junctions that are rotating by an $\times$ as
shown at the top of Fig.~\ref{fig:flux}.  Each branch represents a
junction, but only the ones with a cross are rotating.
Fig.~\ref{fig:flux}(a) shows a type A breather and a plot of the
numerically calculated DC flux per unit cells for a 9 junction ladder
array.  Fig.~\ref{fig:flux}(b) shows the decay of the DC flux for a
type B breather.  The flux decays exponentially so that our breather
solutions are exponentially localized.  In the case of the hybrid
breather shown in Fig.~\ref{fig:1site}(d), the decay of the flux in
the right side of the array is the same as for the type B shown in
Fig.~\ref{fig:flux}(b) and the decay on the left side of the array is
the same as the type A shown in Fig.~\ref{fig:flux}(a).  Note that
we have plotted the modulus of the flux.  The fluxes for all the
cells of one side of the vertical rotating junction have opposite sign
to the fluxes of the cells in the other side.

\begin{figure}[tb]
\epsfxsize=3.2in
\centering{\mbox{\epsfbox{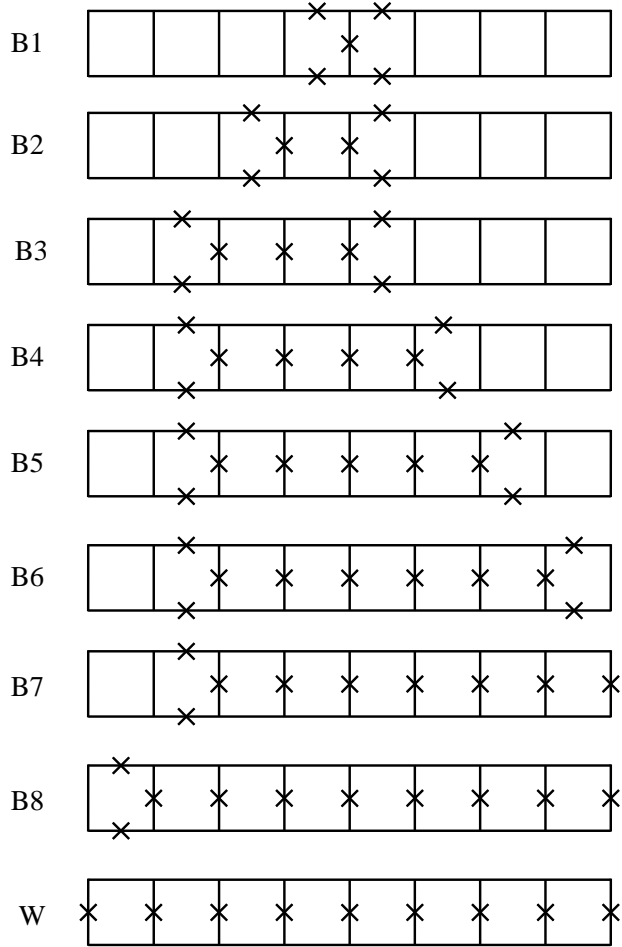}}}
%\vspace{0.15in}
\caption[Schematic of all measured type B breathers]{
Schematic of the DB found
experimentally in {\protect Fig.~\ref{fig:down}}.
Branches with $\times$'s depict rotating
junctions.
B1 through B8 represent type B breathers
with the indicated number of rotating vertical
junctions.  State W is the whirling state
were all of the vertical junction are rotating.
}
\label{fig:ladders}
\end{figure}

Fig.~\ref{fig:ladders} shows different type B breather solutions for which a
set of contiguous vertical junctions are in the rotating state. We call them
multi-site or $m$-site solutions where $m$ refers to the number of vertical
junctions rotating in the array. In general many different solutions are
allowed where each vertical junction can be in the resistive or in the
superconducting state while one or both horizontal junctions between
vertical ones in different states also rotate.

The above discussion focused on the existence of rotobreathers in DC-biased
arrays. In the case of AC-biased arrays the JJ ladder in addition to rotating
localized modes also supports oscillating localized modes or oscillobreathers
where one vertical junction describes a large amplitude oscillation when
comparing with other vertical junctions.  Such modes were studied
in~\cite{floria96,floria98,martinez98} for non-inductive arrays. These modes
persist when inductances are added to the model and also different type-B and
type-A families of AC-biased rotobreathers can be
identified~\cite{unpublished}.

\section{Experiments}
\label{sec:exp}

\subsection{Experimental observation of breathers}

\begin{figure}[tb]
\epsfxsize=\figsize
\centering{\mbox{\epsfbox{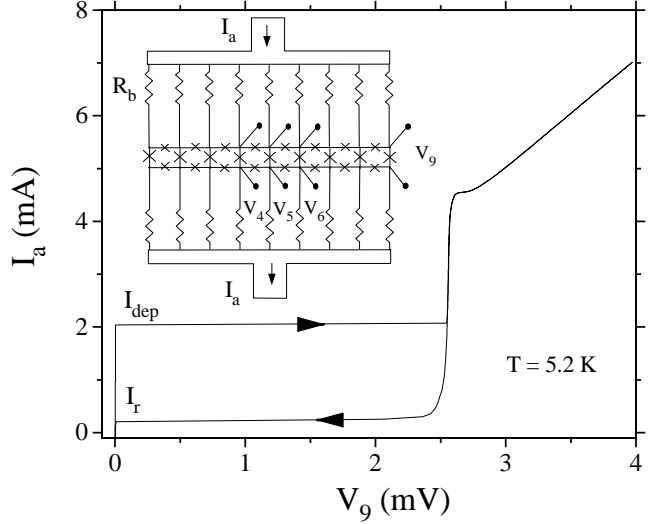}}}
\vspace{0.1in}
\caption[Current-voltage characteristic of an anisotropic ladder]
{Current-voltage characteristic of an anisotropic Josephson
junction array when no breathers are excited.  The hysteresis between
the depinning current ($I_{\rm dep} \approx 2\,{\rm mA}$) and the
retrapping current ($I_r \approx 0.2\,{\rm mA}$) is shown.  Inset:
Schematic of the anisotropic ladder array showing the bias circuit.
$I_a$ is the total applied current and $R_b$ the bias resistances are
$25\Omega$.}
\label{fig:rn}
\end{figure}

We have designed and measured several anisotropic ladders. The inset of
Fig.~\ref{fig:rn} shows a schematic of the measured arrays.
The junctions are fabricated using a Nb-Al$_2$O$_x$-Nb
tri-layer technology with a critical current density of about $1 \, {\rm
kA/cm^2}$~\cite{Hypres}.  The current is injected and extracted
through bias
resistors in order to distribute it as uniformly as
possible in the array.  These resistors are large enough so as to
minimize any deleterious effects on the dynamics.
Our ladders have $3 \times 3\,\mu{\rm m}^2$ 
horizontal junctions and $6 \times 6\,\mu{\rm m}^2$ vertical junctions.
Vertical junctions have been designed with
four times the area of the horizontal ones.  
Thus, the anisotropy ratio $h$ is approximately $0.25$.
The bias resistors are $25\,\Omega$. 

There are voltage probes in the fourth,
fifth, sixth and ninth vertical junctions to measure $V_4$, $V_5$,
$V_6$ and $V_9$. 
The voltage probes can also be used to measure the
top horizontal junctions in the middle, which we denote as $V_{4T}$ and
$V_{5T}$, or any other combination of terminals.

From the measured normal state resistance we calculate $I_{cv}=360 \,
{\rm\mu A}$ and $I_{ch}=90 \, {\rm\mu A}$ at $T$ = 0 K.  The
dimensionless penetration depth $\lambda$, which measures the
inductive coupling in the array, is defined as $\Phi_{0}/2\pi L_s
I_{cv}$.  We estimate the mesh inductance ${L_s}=30.2\,{\rm pH}$ from
numerical modeling~\cite{henry} so $\lambda=0.04$ at $T$ = 0 K.

To determine $\Gamma$ we need to measure the subgap resistance.
Different approaches can be used.  One possibility is to calculate
this damping from the measured return current of the junction.  The
model used for the return current determines the subgap dissipation
and resistance ~\cite{chen88,kirtley88,cristiano97,castellano99}.  We
will estimate the value of $\Gamma$ in our experiments from the
retrapping current $I_r$ by the relation $I_r/N I_{cv} =
(4/\pi)\Gamma$, where $N$ is the number of vertical junctions.  This
expression can be found from a simple energy argument valid for small
values of the damping.  The energy injected into the system by the
applied current must equal the energy to ``rotate'' the junction one
full period~\cite{likharev_book}. Thus from the experiment shown in
Fig.~\ref{fig:rn} we infer a value of $\Gamma \sim 0.08$ ($\beta_c
\sim 160$) at $T=5.2{\rm K}$.

We show a typical IV curve of a ladder without DB in Fig.~\ref{fig:rn}.  We
measure the time-averaged voltage of the 9-th vertical junction as a function
of the uniformly applied current.  The junction is in a zero-voltage state as
we increase the applied current from zero.  When the applied current reaches
the depinning current $I_{\rm dep}$ at about $2\,{\rm mA}$ the junction
switches from zero-voltage state to the superconducting gap voltage, $V_g$,
which at this temperature is $2.5\,{\rm mV}$. If we increase the applied
current further, the junction reaches its normal state and it behaves as a
resistor, $R_n$, of $5\,\Omega$.  As the current decreases the junction
returns to the gap voltage and then to its zero-voltage state at the
retrapping current, $I_r\approx 0.2\,{\rm mA}$.

Sometimes, when we sweep the applied current we find that DB solutions
appear spontaneously: They can be thermally excited when the applied current
is close to $I_{\rm dep}$. However for our experiments, we have developed a
simple reproducible method of exciting a breather: (i) Bias the array
uniformly to a current below depinning current; (ii) increase the current
injected into the middle vertical junction ($V_5$) until its voltage switches
to the gap; (iii) reduce this extra current in the middle junction to zero.
Other procedures are possible.  For instance, we can increase the current for
the middle vertical junction first until it rotates and then increase the
array bias current as was described in~\cite{binder00a}.  We could also inject
this extra current into a horizontal junction.  All of these methods produce
breathers in our arrays and thereby hinting at the generic nature of breather
solutions in our ladders

\begin{figure}[tb]
\epsfxsize=\figsize
\centering{\mbox{\epsfbox{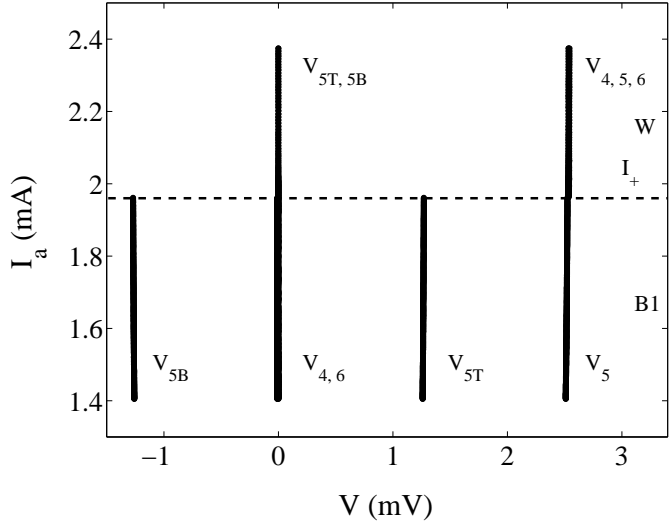}}}
%\vspace{0.15in}
\caption[Breather measurement as current is increased]{Measurement of
the time-averaged voltages of five junctions of the array with the
breather as the applied current is {\it increased} at $T=5.2\,{\rm K}$.  We
first biased the ladder at {1.4\,{\rm mA}} and excited a breather as
indicated in the text.  Then, the applied current is increased.  Below
{$I_+ \approx 1.95\,{\rm mA}$} we see the signature measurement of the
breather and above the breather becomes unstable and the array
switches to the whirling state.}
\label{fig:up}
\end{figure}

Fig.~\ref{fig:up} shows the result after we have excited the breather and we
have increased the array current. The breather is excited at $I_{a} \approx
1.4 \, {\rm mA}$ and then the junction voltages are measured as the applied
current is increased. We simultaneously measure the voltages of the vertical
junctions ($V_4$, $V_5$ and $V_6$) and the top two horizontal junctions,
$V_{4T}$ and $V_{5T}$. The DB is localized in the fifth vertical junction and
is a type B breather since the top, $V_{5T}$, and bottom, $V_{5B}$, horizontal
junctions have opposite voltage.  We also find that the horizontal voltages
are half in magnitude to the rotating vertical junction $V_{5}$.  This is the
B1 breather state shown in Fig.~\ref{fig:ladders}.

The breather exists until a maximum current $I_+ \approx 1.95 \,{\rm mA}$ is
reached. If the applied current is further increased then Fig.~\ref{fig:up}
shows that the voltages of the 4th and 6th vertical junctions jump to the gap
voltage while those of the horizontal junctions go to zero.  In fact all the
vertical junction have jumped to the gap voltage, The ladder has now all of
the vertical junctions rotating and it is in the ``whirling'' state.  The
state is depicted as W in Fig.~\ref{fig:ladders}.  In section~\ref{sec:cir} we
will present a circuit model that will relate $I_{+}$ to the array depinning
current.

\begin{figure}[tb]
\epsfxsize=\figsize
\centering{\mbox{\epsfbox{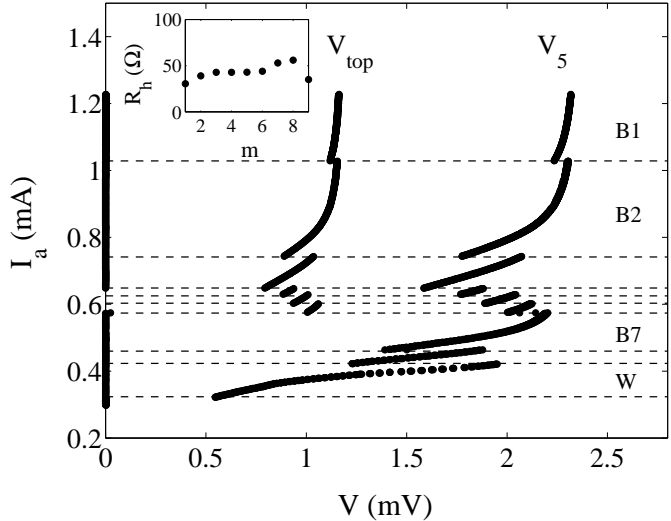}}}
%\vspace{0.15in}
\caption[Breather measurement as current is decreased]{
  Measurement of time-averaged voltages as the applied current is {\it
    decreased} at $T=6.0\,{\rm K}$. We show the voltages of three vertical
  junctions ($V_4$,$V_5$ and $V_6$) and the voltage measured in the top branch
  between the middle and one of the edges of the array ($V_{top}$). We first
  biased the ladder at {1.25\,{\rm mA}} and excited a breather as indicated in
  the text.  Then, the applied current is decreased.  The nine steps
  corresponds to different type B $m$-site breathers.  The dashed lines are
  the expected minimum currents based on a retrapping model [Eq.
  (\ref{eq:minI_Ih})].  The inset shows the $R_h$ values used to fit the
  minimum currents.}
\label{fig:down}
\end{figure}

In Fig.~\ref{fig:down} we measure the voltage as the applied external
current is decreased.  We excite the breather at $I_{a} \approx 1.25
\, {\rm mA}$ and then the voltages are measured as the current is
decreased.  We have measured vertical junctions four, five and six.  We
have also measured the total voltage of the horizontal junctions five
through eight and this sum voltage is referred as $V_{\rm top}$.  This
allows us to reconstruct the $m$-site breather state in terms of the
rotating junctions of the array.

Our experiments show that as we decrease the applied current the single-site
breather will usually decay into an $m$-site breather state.
From $I_{a} \approx 1.25 \, {\rm mA}$ to $I_{a} \approx 1.05 \, {\rm mA}$ the
vertical voltages $V_4=V_6=0$.  This is the single-site type B
breather, the state depicted as B1 in
Fig.~\ref{fig:ladders}.  If we further decrease the applied current from
$I_{-}= 1.05 \, {\rm mA}$, $V_4$ jumps from zero to the gap voltage.  This is
a two site breather shown schematically in Fig.~\ref{fig:ladders} as
B2.  As we
further decrease the current we can count nine 
discontinuous curves, each one corresponding to the switching of
a vertical junction. At {0.3\,{\rm mA}} all of the vertical junctions
return to their zero-voltage state via a retrapping mechanism
analogous to that of a single pendulum.

From these experiments we conclude that this shifting of the voltage
corresponds to at least one vertical junction switching from the
zero-voltage state to the rotating state.  If we assume that only
consecutive junctions switch, then every curve in the measurement can
be associated with one $m$-site breather.  Since we have measured
$V_{\rm top}$ we can reconstruct the ladder solutions.
Fig.~\ref{fig:ladders} shows a schematic of the states measured in
Fig.~\ref{fig:down}.

The shapes of the IV curves in this
multi-site breather regime are influenced by the
junction nonlinear resistance and the redistribution of
current when each vertical junction switches.  This redistribution may
also govern the evolution of the system after each transition to one
of the other possible breather attractors in the phase space of the
array. 

Fig.~\ref{fig:down} shows a measurement at $T=6.0$K and we find 10
different states can be distinguished (8 $m$-site breather states, the
whirling state, and the zero voltage state). In general we only see
three or four $m$-site breathers as we decrease the current, and often
we see different ones even under similar experimental conditions.

\begin{figure}[tb]
\epsfxsize=\figsize
\centering{\mbox{\epsfbox{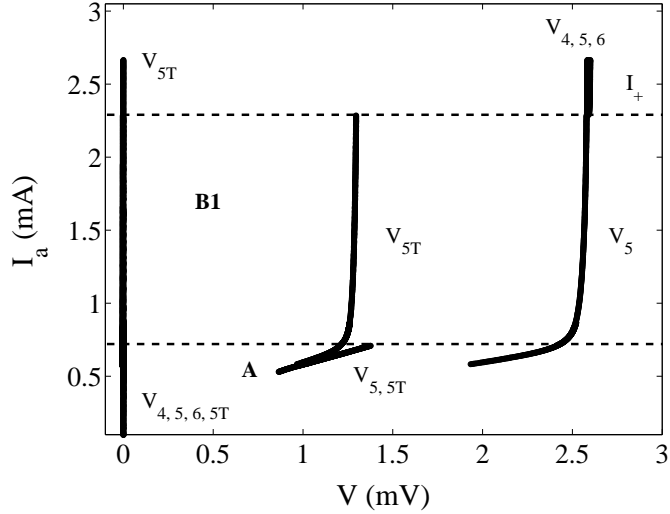}}} 
%\vspace{0.15in}
\caption[Measured asymmetric breather]{
  IV showing type B1 and type A breather at $4.9\,{\rm K}$.  
  Dashed line at $I_a \approx 2.3 \,
  {\rm mA}$ is the maximum current for the type B breather and $I_a \approx
  0.7 \,{\rm mA}$ is the maximum current for the type A breather.  }
\label{fig:asym}
\end{figure}

Fig.~\ref{fig:asym} shows an IV which includes a type A breather.
First, a type B breather is excited at $I_a \approx 0.9 \,{\rm mA}$ and
the current is decreased.  As we decrease the current $V_4$ and $V_6$
have zero voltage while $V_5=2V_{5T}$. At $I_a \approx 0.58 \, {\rm mA}$
the type B breather becomes unstable.  In most of our measurements,
the array will then suddenly switch to an $m$ site breather as shown
in Fig.~\ref{fig:down}, but in this case we find a new switching
behavior: The type B breather switches to a type A solution.  This
state is a type A breather because the voltage $V_5=V_{5T}$ and $V_4$
and $V_6$ are zero.  As the current decreases the type A breather
disappears at $I_a \approx \, 0.53 {\rm mA}$ and the array is in the
superconducting state.  In our experiments, the ladder always returns
to the superconducting state whenever a type A breather reaches its
minimum current.

As the current decreases the type A breather is only accessible for a small
current range of $0.05 \,{\rm mA}$.  However, it is possible to bias on the type A
breather and increase the current to trace out the hysteresis loop.  The
tracing of the type A breather voltage step is also shown in
Fig.~\ref{fig:asym}.  We see that the type A breather exists up to a current
of $\approx 0.7 \,{\rm mA}$.  Once it becomes unstable the array
dynamics usually jumps back to the B1 breather.

\subsection{Temperature and magnetic field dependence}

By sweeping the temperature
and magnetic field we can study how the current range in which our breather
exists is affected by a change of the array parameters.
We define and study the evolution of four current values of importance:
the current when the
array returns to the zero-voltage, $I_r$; the maximum zero-voltage
state current, $I_{\rm dep}$; and the maximum and minimum current for
a breather state, $I_{+}$ and $I_{-}$.  

\begin{figure}[tb]
\epsfxsize=\figsize 
\centering{\mbox{\epsfbox{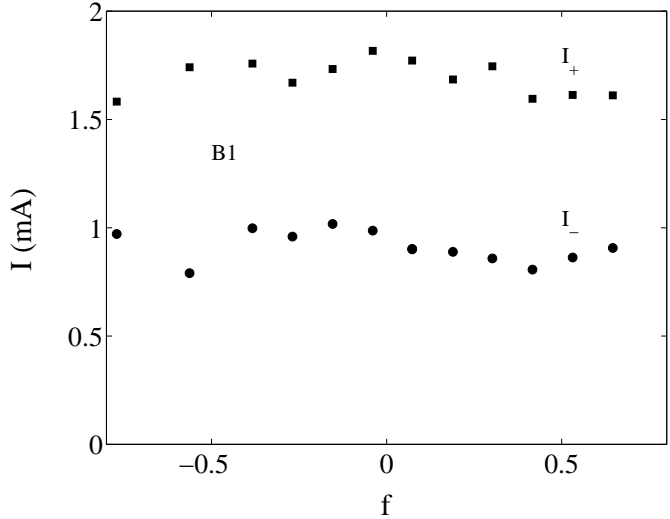}}} 
%\vspace{0.15in}
\caption[Maximum and minimum breather current vs $f$]{
  $I_{+}$ and $I_{-}$ for B1 breather as a function of the applied magnetic field
  at $T=6.0\,{\rm K}$.}
\label{fig:brf}
\end{figure}

\begin{figure}[tb]
\epsfxsize=\figsize 
\centering{\mbox{\epsfbox{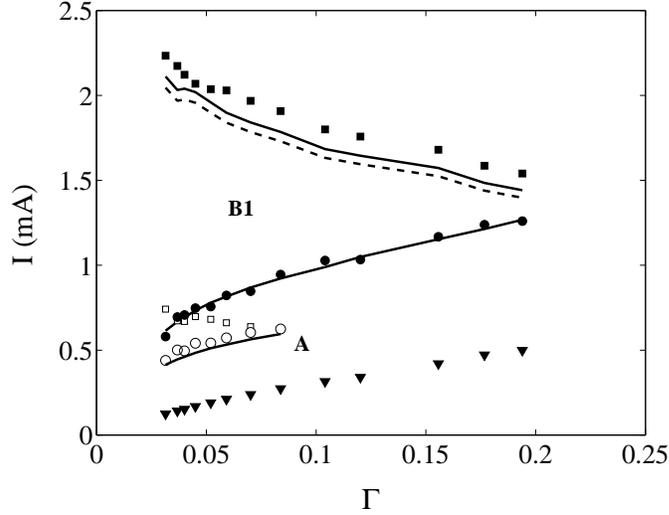}}} 
%\vspace{0.15in}
\caption[Region of existence for type A and B breathers vs $I_{\rm ext}$ and
$\Gamma$]{ Maximum (solid squares) and minimum (solid circles) currents for
  type B breathers, and maximum (open squares) and minimum (open circles)
  currents for type A breathers.  Triangles are the ladder retrapping current.
  Lines through the solid nd open circles are the expected minimum current $I_-$ from
  {\protect Eq. (\ref{eq:minI_Ih})}.  The dashed line is the expected
  uncorrected maximum current from {\protect Eq. (\ref{eq:maxI})} while the
  solid line above it is the corrected value.}
\label{fig:temp}
\end{figure}

Fig.~\ref{fig:brf} shows the typical dependence of $I_{+}$ and $I_{-}$
for a B1 breather as a function of applied magnetic field.  The IV
curves were measured by applying a perpendicular magnetic field of 0
to 300 mG using a magnetic coil that is mounted on the radiation
shield of our probe.  There is some $f$ dependence.  This can be
expected since $I_{+}$ is related to the array depinning current.  But
at this temperature $\lambda=0.06$ so we expect to have a large
Meissner current and a correspondingly relatively flat $I_c$
dependence vs $f$. If $\lambda$ were larger then the breather dynamics
and the $I_c$ should show a stronger magnetic field dependence.

We can study further the existence region of the breather by changing the
temperature of the sample.  In this way, we can vary the parameters to a
certain degree.  The temperature causes the $I_{cv}$ of the junction to change
and hence change $\Gamma$ and $\lambda$.  For our arrays, the junction
parameters can range from $0.031 < \Gamma < 0.61 $ and $0.04 < \lambda <
0.43 $ as the temperature varies from 4.2 K to 9.2 K.

Fig.~\ref{fig:temp} shows how the maximum and minimum current of both type A
and type B breathers are affected as we vary $\Gamma$.  In Fig.~\ref{fig:temp}
$\Gamma < 0.2 $ corresponds to $T < 6.7 \,{\rm K}$ and $\lambda \approx 0.05
$.  At these low temperatures, $I_{cv}$ essentially remains constant so
$\lambda$ does not vary.  However, the subgap resistance varies substantially
as can be seen from the retrapping current measurements in
Fig.~\ref{fig:temp}.  Therefore, there is a larger variation in $\Gamma$.
Figure also shows a nice agreement in between the experiments (points) and the
theory (lines) except for the case of the maximum current of the type A
solutions. Also, experimentally we did not find breathers for $\Gamma>0.2$.

\section{Circuit model}
\label{sec:cir}

\subsection{Introduction}

In this section, we will develop a simplified DC circuit model of our
array in order to understand the region of existence of the breather
solutions in our experiments. Also, this model will allow for
evaluating the effect of the bias resistors in the dynamics of the
array.

In the experiment we apply the external current through bias
resistors as shown in the inset of Fig.~\ref{fig:rn} in order to
distribute it uniformly.  If the resistors are very small, the
horizontal junctions are effectively shunted by a small resistance and
no DB solutions can be excited.  If the
resistors are large so that they dominate over every other
impedance, then applied current will be almost uniform throughout
the array.  However, one drawback of using large bias resistors is
that they will create local heating of the sample and affect the
measurements.

Thus, when studying DB, there are at least two complications
with the bias resistors.  The first is that when we excite a breather state,
the applied current is not completely uniform.
So questions arise about how this non-uniformity affects localization. Also as
we decrease the current we see transitions between different $m$-site breather
states.  In each of these transitions vertical junctions switch from the zero
voltage state to the rotating state. This switching causes a change in the
junction impedance and thereby affects the current distribution.  This
redistribution might be important to understand the dominant drive of the
pattern selection process between $m$-site breathers.  However,
we will show below that the effect of the  bias 
resistors only adds a small correction to the calculation
of the existence region of breather states.

\begin{figure}[tb]
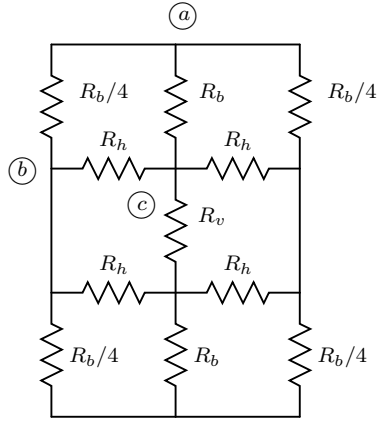

\begin{center}
\input bias
\ \box\graph
\vspace{0.1in}
\caption{Equivalent DC circuit for single-site type B breather
  located at junction 5 in a 9 junction array.  Nodes are labeled a, b, and
  c.}
\label{fig:bias}
\end{center}
\end{figure}

To get some physical intuition we will use a simple model where rotating
vertical junctions have a resistance of $R_v$ and rotating horizontal
junctions have a resistance of $R_h$.  Librating junction will be modeled as
shorts.  We will reduce the array to a simple network of resistors and
calculate DC properties.  The equivalent resistor network for a single-site
symmetric breather located on junction 5 in our 9 junction array is shown in
Fig.~\ref{fig:bias}.

When the array is in the superconducting or in the whirling states,
the applied current distributes uniformly across the vertical
junctions. This will not be the case when we have a breather
since some junctions are in the resistive state while others
are superconducting.  Moreover,
when we have a breather and we measure the voltage of the fifth
vertical junction (Fig.~\ref{fig:down}), we find that its voltage
shifts to a higher value when new vertical junctions switch to the
rotating state. 
These shifts are identified as jumps in the effective
current biasing the junction due to the redistribution of currents in
the array.  Roughly speaking, most of the applied current
wants to flow through the superconducting junctions.
Whenever a vertical junction switches from a superconducting
to a resistive state there is some extra current that
is distributed throughout the array and thereby the effective bias of
the fifth junction becomes a little larger.  This extra bias
results in a jump of the measured voltage to a larger value.
The size of the voltage jump is dependent on the non-linear
subgap resistance and is usually large even for small changes
of the effective bias current.

\subsection{IV curves and current distribution}

In this section we will focus on interpreting this redistribution and shifting
of the voltage 
by using a simple DC model of the ladder.
We use $I_a$ to represent the total applied current in the array.  This is the
current of the experimental current source. The current applied to
a particular junction through the bias resistors, i.e. the current 
through the bias resistor, will be designated $I_j$.  We will use $I_l$ to
denote the current from node a to node b.  This current $I_l$ is the sum of
the currents $I_1$ through $I_4$.  Also, because all of our bias resistors are
the same and due to KVL $I_1=I_2=I_3=I_4$.  So, $I_l=4I_4$ for instance.  The
current through node a to node c is $I_5$ since the breather is located in
junction 5.

For the type B breather we measured,
the voltage is twice as large as the voltage of the
horizontal junction
\begin{equation}
2V_{h}=V_{v}.
\end{equation}
Therefore
\begin{equation}
I_{v}=2 {R_h \over R_v} I_h,
\end{equation}
and we will assume that
$R_v=hR_h$.

The circuit has left-right symmetry
so the current through the right branches must equal the current
through the left. KCL at node $a$ yields
$I_a=2I_l+I_5$ where $I_a$ is the total current applied, and at node $c$
yields $I_5=2I_h+I_v$. KVL on the top left mesh gives $I_l
R_b/4-I_hR_h-I_5R_b=0$.  Combining the equations results in
\begin{equation}
{I_a \over N}= \left \{2+{2 \over h}+\left(1-{1 \over N}\right){R_h \over R_b}
\right \} I_h.
\end{equation}

To generalize to type A breathers we just note that the horizontal junction
voltage is the same as the vertical junction voltage $V_{h}=V_{v}$.  We can
write $s V_{h}=V_{v}$ where $s=1$ for type A breathers and $s=2$ for type B
breathers.  We can also generalize for $m$-site breathers by modifying the
equivalent DC circuit accordingly.  For instance, when $m=2$ both rotating
vertical junctions can be lumped into an equivalent impedance of $R_v/2$.
Generalizing our circuit results in \footnote{These equations are valid for
  type A and type B solutions with $m$ consecutive rotating vertical junctions
  and with junction 1 and $N$ in the no rotating state. In a similar way, we
  can compute the equations for the case when junction 1 or $N$ is rotating,
  for hybrid breathers, or for multi-site breathers for which the rotating
  junctions are not consecutive.}
\begin{equation}
{I_a \over N}= \left \{{2 \over m}+{s \over h}+\left(1-{m \over N}\right){R_h \over R_b}
\right \} I_h.
\label{eq:minI}
\end{equation}

We can also calculate the IV curve of a vertical rotating junction
by substituting $I_h=(h/s)I_v=(h/s)V_v/R_v$ in
Eq. (\ref{eq:minI}).  The result is
\begin{equation}
{I_a \over N} =\left \{1 + {2 h \over s m  } +\left(1-{m \over  N} \right){hR_h \over s R_b} \right \} {V_v \over R_v }.
\label{eq:ciriv}
\end{equation}

Another important variable is the amount of effective
current biasing a vertical rotating junction, say $I_5$,
\begin{equation}
{I_a \over N}= \left \{1+{mh \over 2h+sm}\left(1-{m \over
N}\right){R_h \over R_b}\right \} I_5.
\end{equation}
From our experiments $R_h/R_b \sim 0.8$, $h=0.25$ and $N=9$. Thus for the case
of a type-B single-site breather ($m=1$, $s=2$) we get $I_5=0.934 I_a/9$ and
$I_j=1.008 I_a/9$ ($j \neq 5$). Thus the vertical rotating junction is biased
by a smaller DC current than the quiet vertical junctions. Also this
non uniformity disappears as the bias resistance is made larger.

\subsection{Estimation of $I_{-}$}

To calculate $I_{-}$, we assume that the instability that destroys an $m$-site
breather is due to a junction retrapping mechanism. Even if it appears that
some other mechanism is important, like a resonance, this instability occurs
in the subgap region of the junction where the voltage varies rapidly while
the current remains relatively constant. Therefore, a retrapping current should
give a good estimate of $I_{-}$ regardless of the physical mechanism.

From simple energy consideration of an isolated
junction, the retrapping current can be estimated as $4 \Gamma I_{c} / \pi$.
Then, when the horizontal junction
reaches its retrapping current $I_{h}=4\Gamma I_{ch}/\pi$.
Thus, $I_{v}=4 \Gamma I_{ch} s /\pi h = 4 \Gamma I_{cv}s/\pi$,
and the vertical junction is at $s$ times the junction retrapping
current.  Conversely, if the vertical junction is at
the retrapping current, then the horizontal junction
is at $1/s$ of the junction retrapping
current.  Therefore, as the
applied current decreases,
the horizontal junction always reaches the
retrapping current first for a
type B breather and both the horizontal and
vertical junctions reach the retrapping at the
same time for a type A breather.  This is the reason why
type B breathers can decay into $m$-site breathers
while type A breathers apparently decay into the superconducting state.
When a type A breather reaches $I_-$ all of the rotating junctions
retrap and the resulting state is more likely to be the superconducting
state. On the other hand, when a type B breather reaches $I_-$ only the horizontal
junction retraps while the vertical junctions remain whirling.

From Eq. (\ref{eq:minI}),
\begin{equation}
{I_{-} \over N}= \left \{{2 \over m}+{s \over h}+\left(1-{m \over N}\right){R_h
\over R_b}
\right \} {4 \over \pi}\Gamma I_{ch}
\label{eq:minI2}
\end{equation}
when $I_h$ reaches its retrapping value.

To use this formula we need the horizontal junction parameters.  We
can estimate $C_{h}$ from the specific capacitance of the tri-layer
and the junction area. We find $C_{h}=300 \, {\rm fF}$.  From the
constant $I_cR_n$ product, we find that $I_{ch}=90\,{\rm \mu A}$.
However, the remaining parameter $R_h$ is more difficult to estimate
since it depends highly on the nonlinear subgap region.  Instead of
trying to calculate $R_h$ directly, we will fit Eq. (\ref{eq:minI2})
to our measured $I_{-}$ using $R_h$ as our fitting parameter.  We
include the effect of $R_h$ through the definition of
$\Gamma=\sqrt{\Phi_0/2\pi I_{ch} R_{h}^2C_{h}}$.  Then from
Eq. (\ref{eq:minI2}),
\begin{equation}
{I_{-} \over N}= \left \{{2 \over m}+{s \over h}+\left(1-{m \over N}\right){R_h \over R_b}
\right \} 
{4 \over \pi}\sqrt{{\Phi_0 \over 2\pi R_{h}^2C_{h}}I_{ch}}
\label{eq:minI_Ih}
\end{equation}
and $R_h$ as the only free parameter.  
This is how we
estimate $R_h$, or equivalently $\Gamma$, for a given temperature.

The inset of Fig.~\ref{fig:down} shows the fitted values for $R_h$ and
the dashed lines in Fig.~\ref{fig:down} show the resulting $I_{-}$.  We note
that for solutions B7, B8, and $W$ we used a different equivalent circuit
based on the schematic of  Fig.~\ref{fig:ladders}.  
We see that $R_h$ is approximately $50 \,
\Omega$ for all $m$.  This value is not totally unexpected.
Roughly speaking, the horizontal junctions are
shunted by two bias resistor of $25 \, \Omega$ each and the retrapping
current depends strongly on the equivalent junction impedance.  Since the
subgap resistance for our junctions can be several ${\rm k}\Omega$ the
equivalent horizontal junction impedance will be dominated by the shunts
which add up to $50 \, \Omega$.

We can also easily understand why $I_{-}$ decreases as $m$ increases.  For a
constant array bias at $m=1$ some fraction of the current will flow through
both horizontal junctions.  At the same bias current for $m=2$, there is a
horizontal junction that is not rotating in between the rotating horizontal
junctions.  By symmetry considerations there is no current flowing through
this quiet horizontal junction.  Since the applied bias current is the same,
this implies that a larger fraction of the applied current must flow through
the rotating horizontal junctions when $m=2$ than when $m=1$.  Thereby, $I_{-}$
is smaller for $m=2$. 

\subsection{Estimation of $I_{+}$}

To calculate $I_{+}$ we look for the librating junction that supports the
maximum current.  When this junction reaches $I_c$, the breather has reached
its maximum current.  It is straightforward to find that the critical junction
is the first vertical junction that is not rotating (the one nearest to the
rotating ones). Let $I^\star$ be the current through the junction. KCL at node
b of the array yields $I^\star=I_h+I_4$.  Here we have assumed that there is
not current in the horizontal quiet junctions.  Since $I_l=4I_4$ we can
substitute for the currents to solve for $I^\star$ in terms of $I_a$.  We can
also generalize to $m$-site breathers. The result is
\begin{equation}
{I_a \over  N}={2h/m+s+h\left (1-m/N\right )R_h/R_b
\over h(1+2/m)+s+hR_h/R_b}I^\star.
\label{eq:maxI}
\end{equation}
The maximum applied current $I_+$ occurs when 
$I^\star=I_{cv}$.
In the limit $R_b\gg R_h$,
\begin{equation}
{I_{+} \over N}={2h+sm \over h(m+2)+sm}I_{cv}.
\label{eq:i+}
\end{equation}

This current will underestimate the actual value for $I_{+}$.  This is because
we have not taken into account any of the horizontal junction currents that
are in the quiet state. 

The effect of fluxoid quantization is to redistribute the currents of the
quiet junctions.  The currents in the bias resistors and the rotating
junctions will remain unaffected. If we consider the effect of the next
nearest mesh to the breather, then KCL at node b will be $I_4 + I_h =I^\star +
I_{ch}\sin(k_h)$.  Here, $I_{ch}\sin(k_h)$ is the current in the next
horizontal junction.  Fluxoid quantization in this quiet mesh yields
$k^\star-k_v-2k_h=0$ when $f=0$ and we neglect the self induced field.  Here
$k^\star=\sin^{-1}(I^\star/I_{cv})$ and $k_v$ is the phase of the next quite
vertical junction.  Adding self fields will tend to decrease $I_{ch}\sin(k_h)$
because smaller screening currents are needed as the inductance becomes
larger.  This correction will then tend to overestimate $I_{+}$.  We note
that to calculate $I_{+}$ we set $I^\star=I_{cv}$ so $k^\star=\pi/2$.

With fluxoid quantization $k_h= \pi/4 -k_v/2$ at $I_a=I_{+}$.  To
first order we expect the current in the quiet vertical junction to be
$I_a/N$ and $k_v=\sin^{-1}(I_a/NI_{cv})$.  For consistency, we apply
KCL so $I_3+I_{ch}\sin(k_h)=I_{cv}\sin(k_v)$.  Again, we neglect the
current of the next quiet horizontal junctions.  We can solve for
$k_v$ in terms of $k_h$ and substitute back into the fluxoid
quantization condition.  Using $\sin x\approx x$ and $\cos x \approx
1-x^2/2$ we can get a closed expression
\begin{equation}
k_h={-h+\sqrt{h^2+8(1-I_3/I_{cv})} \over 4}.
\label{eq:kh}
\end{equation}
This expression is only valid when $I_{a}\approx I_{+}$.

The current $I_3$ equals $I_4=I_l/4$ and can be calculated from
Fig.~\ref{fig:bias} and is simply
\begin{equation}
I_3={2/m+s/h+R_h/R_b \over 2/m+s/h+(1-m/N)R_h/R_b}{I_a \over N}.
\label{eq:I3}
\end{equation}
In the limit $R_b \gg R_h$ $I_3=I_a/N$.  

The maximum current will
now be Eq. (\ref{eq:maxI}) when $I^\star=I_{cv}+I_{ch}\sin(k_h)$ with $k_h$
defined in Eq. (\ref{eq:kh}) and Eq. (\ref{eq:I3}).  The resulting equation is
transcendental.  But we know the correction due to $k_h$ should be small so we
can linearize the sine term.  The equation then has dependencies of $I_a$ on both
sides but can be solved by isolating the square root and squaring.  This
results in a quadratic equation in $I_a$ with easily extractable roots.  The
equation can also be solved iteratively.  One iteration usually results in a
good approximation.  Hence, if $I_{+}^0$ is the uncorrected value from
Eq. (\ref{eq:maxI}) when $I^\star=I_{cv}$, then the first correction $I_{+}^1$
can be calculated when $I^\star=I_{cv}+I_{ch}\sin(k_h)$ with $I_a=I_{+}^0$
substituted in in Eq. (\ref{eq:I3}).

\subsection{Comparison with the experiments}

The result of the simplified DC circuit model is compared to the
measurements in Fig.~\ref{fig:down} and Fig.~\ref{fig:temp}.  We used
the measurement in Fig.~\ref{fig:down} to estimate the values of $R_h$
and we use Eq. (\ref{eq:minI_Ih}) to calculate $I_{-}$ for the type A
and B breathers.  In Fig.~\ref{fig:temp}, we see that the result
agrees quite well for $I_{-}$.  We have also calculated $I_{+}$ from
our uncorrected expression Eq. (\ref{eq:maxI}) and with the correction
due to Meissner currents.  The measured and expected results agree for
type B breathers.  However, the measured $I_{+}$ for the type A
breathers is much lower than what our circuit model predicts.

We also note that the effect of the bias resistors and $R_h$ is
essentially to provide an offset to the quasi-linear slope in
Fig.~\ref{fig:temp}.  In the limit $R_b \gg R_h$, the
predicted values for $I_{-}$ would intersect ($0,0$).  The bias
resistor, then, only provide a small correction by allowing a better
fit to the measurements.

Finally, there appears to be a maximum amount of damping where breathers can
exist in the ladder.  This maximum $\Gamma$ occurs when $I_{+}$ coincides with
$I_{-}$ which occurs at $\approx0.2$.  This can be expected from the fact that
the DB discussed in this paper require hysteresis in the IV curves.

\section{Simulations at the experimental values}
\label{sec:brsims}

\subsection{Introduction}

In section~\ref{sec:ce} we derived the standard model for the dynamics
of the array. This model assumes the RCSJ for the junction dynamics,
uniform bias condition, and the effect of mesh self-inductances.  We
will use numerical simulations of the governing equations to provide
for detailed analysis of the DB solutions.  We numerically integrate
Eq. (\ref{eq:ladder}) using a 4th order Runge-Kutta method.  The
initial condition is found via a similar procedure as in the
experiments.  We first bias the array near the depinning current.
Then we increase the current of the middle vertical junction until it
starts to rotate and finally we decrease this extra current to zero.
Usually the resulting solution is periodic and we verify that it is
stable by calculating Floquet multipliers. If the solution is not
stable we integrate the equations of motion for a long time with an
added small current noise source in every junction. In this way we
perturb the solution and sample nearby trajectories in phase
space. Usually the final result is a new periodic stable solution.

Sometimes, and usually close to the destabilization points, when we
integrate the equations we find aperiodic solutions that appear
stable.  We again add some noise to check the stability of the
solution against fluctuations and use standard methods like Poincar\'e
section analysis or study of the Liapunov exponents of the system to
gain more information about the behavior of the solution.

\subsection{Numerically integrated IV curves with a nonlinear resistance}
\label{sub:nlr}

We will present numerical simulations of the dynamics of the
ladder. In our samples $h=0.25$ and the damping and the penetration
depth vary with the temperature. Thus when the temperature varies from
4.2 to 9.2 K, $\Gamma$ varies from 0.03 to 0.6 and $\lambda$ from 0.04
to 0.4. However, experimentally we only find breathers for $T<6.7$K
and that corresponds to $0.03 < \Gamma < 0.2$ and $0.04 <\lambda <
0.05$.  We will then mostly present simulations at $h=0.25$,
$\lambda=0.05$, however the dynamics of the array is very rich and
multiple transitions between different attractors occurs when changing
the parameters.

\begin{figure}[tb]
\epsfxsize=\figsize
\centering{\mbox{\epsfbox{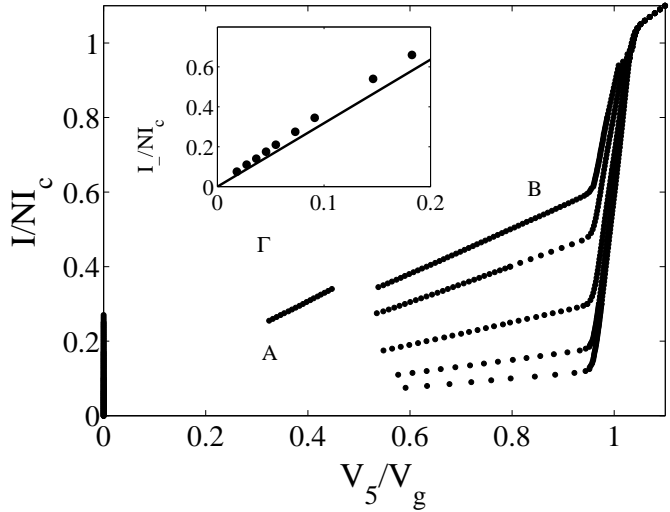}}}
%\vspace{0.15in}
\caption[Simulation of type B breather] {Simulation of type B breather
with a subgap resistance modeled by Eq.  (\ref{eq:subgapmodel}).  From
bottom to top $g_{sg}=0.1, 0.15, 0.25, 0.4,$ and $0.5$.  $\Gamma=0.18$
and $\lambda=0.05$.  When
decreasing the current the type B breather destabilizes to a type A
when $g_{sg}=0.5$.
Inset shows the plot of $I_{-}$.  Solid circles
are from simulations and the line is from Eq.  (\ref{eq:minI2}). 
}
\label{fig:LadNCSims}
\end{figure}

Fig.~\ref{fig:LadNCSims} shows simulated IV curves of single-site type
B breather solutions.  These simulations were done with a fixed
$\Gamma=0.18$ and $\lambda=0.05$.  We have included a subgap
resistance in this simulation in order to compare with the
experimental measurements more closely.  We use the usual approach to
extend the RCSJ model where now $R$ depends on the junction
instantaneous voltage $V(t)$.  We define a conductance such that
\begin{equation}
G(V(t))=\left \{ \begin{array}{ll}
              {1 \over R_{sg}(T)} & \mbox{if $|V(t)|<V_g$} \\
              {1 \over R_n} & \mbox{otherwise}
              \end{array}
     \right .
\end{equation}
Here $R_{sg}(T)$ is taken to be only a function of temperature so for
a given set of parameters it is constant.  Our functional ${\cal
N}(\varphi)$ now becomes $\ddot{\varphi}+ \Gamma
g(v)\dot{\varphi}+\sin\varphi$, where $g(v)=R_n/R_{sg}(T)$ and
$\Gamma$ is calculated from $R_n$.

A simple approach is to model $g$ as a continuous hyperbolic tangent.  We will
use
\begin{equation}
g(v)= g_{sg}+(1-g_{gsg}){1-\tanh K(1-v) \over 2}.
\label{eq:subgapmodel}
\end{equation}
In our simulation,
we take the value of $K=100$.  Thereby
Eq. (\ref{eq:subgapmodel})
approaches a piecewise linear function with a conductance
$g_{sg}$ at $v<1$ and 1 when $v>1$.

We find that the simulated curves are similar to the experimentally
measured arrays.  The inset compares $I_{-}$ of the simulation to the
prediction of the circuit model Eq. (\ref{eq:minI2}) when
$R_b \gg R_h$.  The deviation is due to inductance effects.
Our simple circuit model neglects the effects of the inductances.  We
have also verified that $I_{+}$ in the simulations is within 99\% of
the adjusted $I_{+}$ of our model Eq.  (\ref{eq:maxI}) and
Eq. (\ref{eq:I3}).  We have found similar results for numerically
computed type A breathers.

There is little effect of the subgap resistance in our simulations
besides changing the shape of the IV curve.  Instead of using a subgap
resistance, we can redefine $\Gamma$ so that it includes the effects
of $g$.  Then, $\Gamma$ is calculated from $R_{sg}$ and ${\cal
N}(\varphi)=\ddot{\varphi}+ \Gamma\dot{\varphi}+\sin\varphi$ as
before.  

\subsection{Breather instabilities and switching mechanism}

\begin{figure}[tb]
\epsfxsize=5.0in
\centering{\mbox{\epsfbox{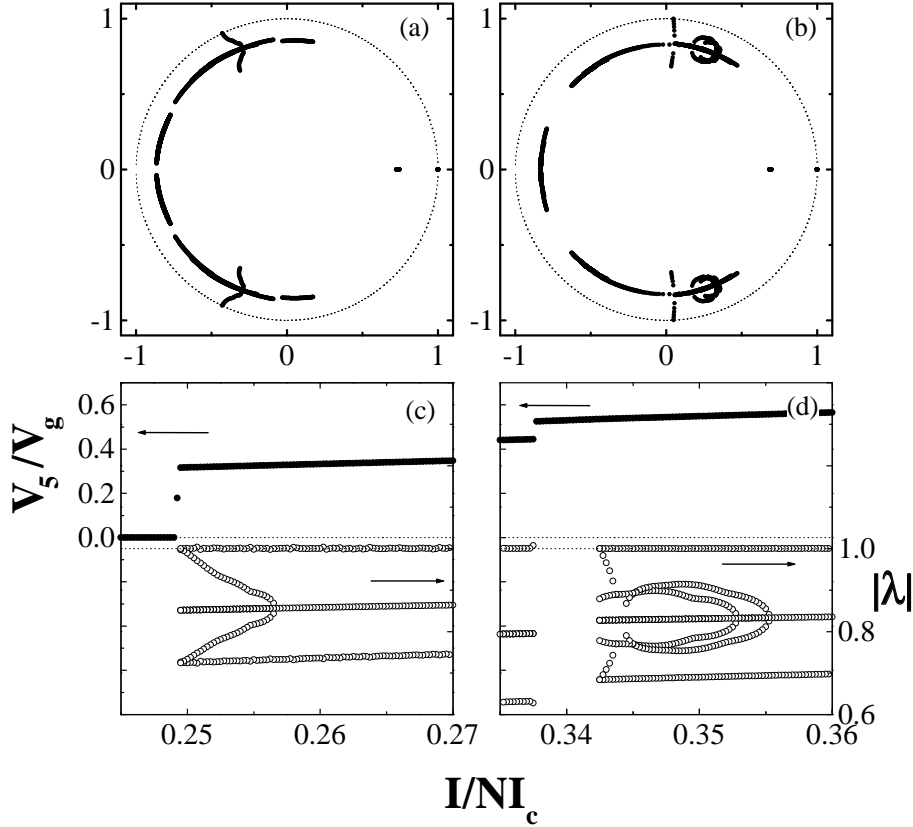}}}
%\vspace{0.15in}
\caption[Floquet multipliers at $I_{-}$ of type B breather] {Floquet
  multipliers of type A (a) and type B (b) periodic DB for decreasing currents
  above and at $I_{-}$ of the simulation shown in Fig.~\ref{fig:LadNCSims}
  when $g_{sg}=0.5$. Figures (c) and (d) show as a function of the current the
  value of voltage (solid circles) and the modulus of the Floquet multipliers
  whenever the solution is periodic (open circles).}
\label{fig:LadNCFq}
\end{figure}

Another aspect of Fig.~\ref{fig:LadNCSims} is the resulting state of the array
once the single-site type B breather becomes unstable.  Some insight can be
gained by first studying how the single-site breather destabilizes.
Fig.~\ref{fig:LadNCFq}(b) shows the distribution of the Floquet multipliers at
several values of the applied current close to $I_{-}$ for the type B breather
when $g_{sg}=0.5$ in Fig.~\ref{fig:LadNCSims}.  We use $I_{\rm ext}=0.36,
0.35975, \ldots, 0.3425$.  The voltage of the vertical junction $V_5$ is
approximately $0.55 \, V_g$ in this small range of currents.  Correspondingly,
the voltage of the top horizontal junction is $V_{5T}=V_5/2=0.275 \, V_g$.  In
our normalization, the fundamental frequency of the type B breather is then
$\omega=V_{5T}/\Gamma=1.53$.  Most multipliers lie on a circle of radius
$e^{-\pi\Gamma g_{sg}/\omega}\approx 0.83$ and this can be verified in
Fig.~\ref{fig:LadNCFq}(b).

In Fig.~\ref{fig:LadNCFq}(d) we decrease the current and show the value of
the voltage and the modulus of the Floquet multipliers (only for periodic
solutions). This is the typically observed bifurcation scenario for small
$\lambda$.  It seems that for small $\lambda$ and underdamped junctions, this
instability introduces more frequencies in the solution.  When the periodic
type B breather losses stability at $I=0.3425$, the solution becomes a
quasi-periodic type B breather similar to the one shown in
Fig.~\ref{fig:quasi}.  This quasi-periodic type B solution persists up to a
current value of 0.33775 when the array jumps to a periodic type A solution.

For large $\lambda$, however, we usually observe a period-doubling bifurcation
where a multiplier crosses the unit circle at $-1$, though the behavior also
depends on the damping.

Our simulations in Fig.~\ref{fig:LadNCSims} show that the array sometimes
switches to the superconducting state when the type B solution ceases
to exist. The bifurcation scenario for the different subgap
resistances is similar to Fig.~\ref{fig:LadNCFq}(b).

At $g_{sg}=0.5$, on the other hand, when the single-site type B breather
becomes unstable, the array switches to a single-site A breather. This type A
breather exists for a range of currents.  Fig.~\ref{fig:LadNCFq}(a) shows the
Floquet multipliers for a type A breather at current values: $0.27, 0.26975,
\ldots, 0.2495$.  The voltage of the vertical junction $V_5$ is approximately
$0.33\,V_g$ in this range of current.  In our normalization, the fundamental
frequency of the type A breather is then $\omega=V_{5}/\Gamma=1.81$.  Most
multipliers lie on a circle of radius $e^{-\pi\Gamma g_{sg}/\omega}\approx
0.85$ and this can be verified in Fig.~\ref{fig:LadNCFq}(a).

In Fig.~\ref{fig:LadNCFq}(c) we decrease the current and show the value of
the voltage and the modulus of the Floquet multipliers (only for periodic
solutions). Below $I=0.2495$ the DB is unstable and the solution switches to
the superconducting state.

The selection rules between different $m$-site type B solutions shown by the
experiments is an important feature that can not be explained in the framework
of the DC circuit model and also is not well predicted by our simulations of
the array.  Once the breather solution reaches a critical value of the current
the system chooses between many different attractors that have complex
boundaries.  In this process randomness and thermal fluctuations play a role
as it is shown by the different switching patterns under similar experimental
conditions. In our experiments we usually observe transitions from a $m$-site
type B solution to a $m+n$-site type B. However sometimes we find jumps from
type B to type A in the experiments (Fig.~\ref{fig:asym}) The simulations
(like in Fig.~\ref{fig:LadNCSims}) show more frequently transitions from a
type B solution to a type A solution and then, by decreasing the current, to
the superconducting state.

To try to understand this switching process of the cascade of $m$-site type B
breathers, we have introduced more elaborated models to (i) include the bias
circuit in our simulations, (ii) study the effect of external fields, (iii)
introduce a full-inductance matrix formalism, (iv) take into account the
nonlinear character of the junction resistance, (v) include disorder
randomizing the junctions critical currents and (vi) include thermal effects
by adding a noise term to the junction currents. However, none of these
simulations reliably predicts the observed cascade of $m$-site type B
breathers found in our experiments as the current is decreased.

The bias resistors can be added to the model by rewriting the KVL and
KCL for the new circuit. We have also included an external magnetic field.
However at small values of $\lambda$ the field is expelled from the
array because of strong inductive effects and does not affect the IV
curves.  This was also found experimentally in Fig.~\ref{fig:brf}.

It also relatively straightforward to use a full-inductance matrix.  We find
that the additions of extra coupling changes significantly the decay of the
fields within the array. For instance, due to the non-local coupling the
breathers are no longer exponentially localized, and the field decay is
algebraic away from the breather.  Nonetheless, the IV remains relatively
unaltered.

Finally, thermal effects may play an important role, especially if the
attractors have complex boundaries.  We use the standard Langevin
approach and replace the resistor of the RCSJ model by a noiseless
resistor in parallel with a Johnson current noise source,
\begin{equation}
C_j\dot{v}_j+{v_j \over R_j}+I_c^j\sin\varphi_j=I_{\rm ext}+I^N_j
\end{equation}
where $<I^N_j(t)I^N_k(t')>=(2kT/R_j)\delta(t-t')\delta_{jk}$.
This results in the usual current noise spectrum density
$S_j=2kT/R_j$.  We normalize our equations
as in Sec.~\ref{sec:ce} and 
${\cal N}(\varphi_j)= \ddot{\varphi}_j+\Gamma
\dot{\varphi}_j+\sin{\varphi_j}+i^n_j$ and
the spectrum
of $i^n_j$ is $S_j=2kTh_j\Gamma/E_J$ where the Josephson
energy $E_J=(\Phi_0/2\pi)I_{cv}$ and
$h_j=I_c^j/I_{cv}$.

Our dimensionless temperature is then $\tilde{T}=kT/E_J$.  At $4.2\,
{\rm K}$ our vertical junctions have $I_{cv}= 345\,{\rm \mu A}$ so
that $E_J=8.2\times 10^3 \, {\rm K}$.  With this normalization $4.2\,
{\rm K}$ equals $\tilde{T}\approx 5.1\times 10^{-4}$.  Similarly, for
the maximum temperature we observe breathers, $6.7\, {\rm K}$,
$\tilde{T}\approx 1.3\times 10^{-3}$.

Our simulations were done at $h=0.25$ and $\lambda=0.05$. At these
values of the parameters, we generally find in our simulations with
Johnson noise that when a type B breather becomes unstable it jumps to
a type A breather.  Sometimes when a type B breather becomes unstable
an $m$-site B breather will form as in our experiments, but this is
much less common. However, we have found that at other values of the
parameters (typically larger $h$ and $\lambda$) different switching
scenarios occur.  We have run several simulation that include thermal
noise, the full inductance matrix, bias resistors, and possible stray
fields.  All these effects result in a very detailed model for the
dynamics of the array and different breather solutions are
found. Also, adding inhomogeneities, like a distribution of junction
critical currents, or a magnetic field does not seem to change the
fact that in the array simulations when most type B breather solutions
become unstable a type A breather solution is formed.

To investigate the experimentally observed cascade of $m$-site type B
solutions we study a toy model that allows for only type-B breathers.
The simplest model with this characteristic is a variation
of the standard model where an up-down symmetry for the
superconducting phases is assumed, thus
$\varphi^t_j(t)=-\varphi^b_j(t)$ (this was also assumed
in~\cite{flach99} and~\cite{giles00}). We remove the bottom branches
of the full ladder, since their dynamic is by definition identical to
the top horizontal junctions.  Also, fluxoid quantization,
Eq. (\ref{eq:fqbrlad}), is modified since $\varphi^t_j=-\varphi^b_j$,
\begin{equation}
\varphi^v_j-\varphi^v_{j+1}-2\varphi^t_j+2\pi f+{1\over \lambda}i^m_j =0.
\end{equation}

This along with KCL yields a nonlinear coupled system that can be
obtained from Eq. (\ref{eq:ladder}) defining $\varphi^b_j=0$ and
changing $\varphi^t_{j}\rightarrow 2\varphi^t_{j}$ on the right hand
sides.

\begin{figure}[tb]
\epsfxsize=\figsize
\centering{\mbox{\epsfbox{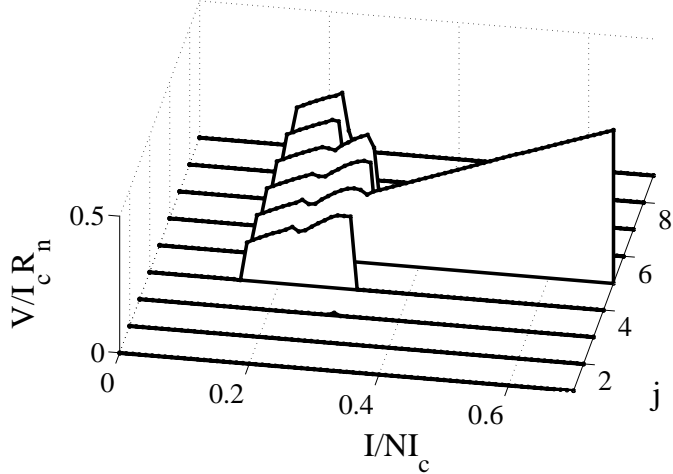}}}
%\vspace{0.15in}
\caption[Simulation of breather solution in up-down symmetric
ladder]{Simulation of a cascade of 1-site, 4-site, and 6-site type B
breathers in up-down symmetric
ladder with $\Gamma=0.07$ and $\lambda=0.05$. Each
curve is the average of the indicated vertical junction.
}
\label{fig:pltHalfL}
\end{figure}

Fig.~\ref{fig:pltHalfL} shows a simulation of the up-down symmetric
ladder with $\Gamma=0.07$ and $\lambda=0.05$.  The curves are the
average voltages of each vertical junctions.  We start at $I_{\rm
ext}=0.7$ with a single rotating junction.  As we decrease the
current, the single-site breather becomes unstable at $I_{\rm
ext}=0.32$ and a 4-site breather is created.  This solution persists until
$I_{\rm ext}=0.23$ when a 6-site breather is formed and finally at
$I_{\rm ext}=0.15$ the arrays switches back to the zero-voltage state.
These current values are estimated well from our circuit model
Eq. (\ref{eq:minI2}) and the discrepancies are due to the effect of
the inductances.

The up-down symmetric ladder only allows up-down symmetric solutions
such as a type B breather.  When the single-site breathers becomes
unstable the array can no longer jump to an A type breather.  It
appears that this constraint is enough to allow the formation of
$m$-site breather when the single-site breather becomes unstable.
This toy model, thus allows for the study of the switching seen in the
experiments, though it is not clear the physical relation of this
model to the experiments.

\section{Linear Analysis}
\label{sec:ta}

\subsection{Resonances in a Ladder}

Before we embark on an analytical study of the different localized
solutions in the ladder we first need to understand the basic
linearized excitations that can occur.

An important characteristic frequency of the ladder occurs when the
frequency of a junction resonates with the lattice eigenmodes.  To
calculate the resonant frequencies, we linearize Eq. (\ref{eq:ladder})
around a solution.  Every breather solution is approximately
up-down symmetric far from the rotating junctions.  Therefore we make
the approximation $\varphi^t_j=-\varphi^b_j$ and let
$\varphi^t_j=\varphi^t_0+\delta\varphi^t_j$ and
$\varphi^v_j=\varphi^v_0+\delta\varphi^v_j$.  The resulting linear
equations are
\begin{eqnarray}
h(\delta\ddot{\varphi}^t_j + \Gamma\delta\dot{\varphi}^t_j+\cos(\varphi^t_0)\delta\varphi^t_j) 
&=&\lambda(-\delta\varphi^v_{j+1}+\delta\varphi^v_j-2\delta\varphi^t_j)  
\label{eq:lin_ladder} \\
\delta\ddot{\varphi}^v_j + \Gamma\delta\dot{\varphi}^v_j+\cos(\varphi^v_0)\delta\varphi^v_j 
&=&\lambda(\delta\varphi^v_{j+1}-2\delta\varphi^v_j+\delta\varphi^v_{j-1}
+2\delta\varphi^t_j-2\delta\varphi^t_{j-1}) \nonumber
\label{eq:lin_ladder2} 
\end{eqnarray}

If the ladder is in the uniform whirling state then the
approximate solution is $\varphi^t_0 = 0$ and
$\varphi^v_0 =\omega t + z j$ where $z$ is the 
wavelength of the vortex train.  To calculate the dispersion relation we
let $\delta\varphi^t_j=\epsilon^t e^{i(zj+\omega t)}$ and
$\delta\varphi^v_j=\epsilon^v e^{i(zj+\omega t)}$.  We substitute in Eq.
(\ref{eq:lin_ladder}) and take the $\Gamma=0$ limit since our junctions are
underdamped and it can be shown that for $\Gamma<1$, the damping terms are
only small corrections to the frequency.  The
$\cos(\varphi^v_0)\delta\varphi^v_j$ results in terms of with coefficients of
$e^{i2\omega t}$ and $e^{0}$ and all other terms have coefficients of
$e^{i\omega t}$.  We only keep terms of $e^{i\omega t}$.  The resulting
matrix equation is
\begin{equation}
\left[
\begin{array}{cc}
-\omega^2+1+{2 \lambda \over h} & {\lambda \over h}(e^{iz}-1) \\
2\lambda (e^{-iz}-1) & -\omega^2+4\lambda\sin^2({z \over 2})
\end{array}
\right]
\left[
\begin{array}{cc}
\epsilon_t \\
\epsilon_v
\end{array}
\right]
=
\left[
\begin{array}{cc}
0 \\
0
\end{array}
\right].
\label{eq:ladmat1}
\end{equation}
The solution to Eq. (\ref{eq:ladmat1}) is $\omega^2=F\pm\sqrt{F^2-G}$ where
$F=(1+2\lambda/h+4\lambda\sin(z/2)^2)/2$ and $G=4\lambda\sin(z/2)^2$.  From
physical grounds we expect the wavelength to be well approximated by $z=2\pi
f$, i.e. the average distribution of vortices in the array.  
The resulting resonant frequencies are important when studying properties of 
{\em moving} vortices in the ladder.  Then, there is a  traveling wave
of vortices with density $f$ that can resonate with the
lattice modes.

Our breathers solutions are clearly not uniform.  So the
above result, while instructive, is of limited value.  We can think instead of
a solution that is valid far from the localized breather core. 
Far from the breather core, the
solution to first order is 
\begin{eqnarray}
\varphi^t_0 &=& \pi f  \nonumber \\
\varphi^v_0 &=&\sin^{-1}i_{\rm ext}
\end{eqnarray}
and this satisfies the ladder equations, Eq. (\ref{eq:ladder}).  
Since the core junctions
in the breather are rotating, they will induce librations in all of the
junctions in the array~\cite{flach99}.  
After substituting the first order solution into Eq. (\ref{eq:lin_ladder}) 
and expanding the perturbations as plane waves,
we are left with the
following matrix that must have a zero determinant to allow for nontrivial
solutions:
\begin{equation}
\left[
\begin{array}{cc}
-\omega^2+i\Gamma\omega+\cos(\pi f)+{2 \lambda \over h} & {\lambda \over h}(e^{iz}-1) \\
2\lambda (e^{-iz}-1) & -\omega^2+i\Gamma\omega+p+4\lambda\sin^2({z \over 2})
\end{array}
\right]
\left[
\begin{array}{cc}
\epsilon_t \\
\epsilon_v
\end{array}
\right]
=
\left[
\begin{array}{cc}
0 \\
0
\end{array}
\right]
\label{eq:resmat}
\end{equation}
where $p=\sqrt{1-(i_{\rm ext})^2}$.  We set $f=0$ as in the 
simulations and $z$ represents the wavelength of the
perturbations.  This dispersion relation describes the 
linearized frequencies that can resonate with a breather.
In the case where $p$ is approximately 1 and $\Gamma=0$ 
the eigenvalues have a particular
simple  form
\begin{eqnarray}
\omega_{+}  = & \sqrt{1+{2\lambda \over h}+4\lambda\sin^2\left({z/2}\right)} \nonumber \\
\omega_{-}  = & 1
\label{eq:res}
\end{eqnarray}
we note that $\omega_{+}>\omega_{-}$ and that $\omega_{-}$
is associated with the $L_J C$ resonance of the ladder
and $\omega_{+}$ with the $L_s C$ resonance.
We will show that this is a good approximation for our simulations.

A more generalized resonance condition can be calculated by using a harmonic
balance technique.
For the added complexity of the harmonic balance analysis, it only
yields a result that is very similar to the linearized calculations.
However, the harmonic
balance approach can also be used to calculate the IV curve.

\subsection{Decay length}

To calculate a resonance frequency we substitute into (\ref{eq:res}) the
appropriate wavelength.  For instance, the resonance of largest frequency
occurs when $z=\pi$.  However, whenever our frequency or wavelength falls
outside of the dispersion relation then any linearized excitations must decay.
In general we should let our wavelength be a complex number $z=k+i\xi$ where
$\xi$ represents the decay length.

We first define $\sin^2(z/2)=x+iy$ and substitute into the
determinant of matrix Eq. (\ref{eq:resmat}).  We expand 
terms and solve for $x$ and $y$,
\begin{equation}
\left[
\begin{array}{cc}
 4\lambda(\omega^2-1) & 4\Gamma\lambda\omega \\
4\Gamma\lambda\omega & -4\lambda(\omega^2-1)
\end{array}
\right]
\left[
\begin{array}{c}
x \\
y
\end{array}
\right]
=
\left[
\begin{array}{c}
\omega^4-(1+p+2{\lambda \over h}+\Gamma^2)\omega^2+2{\lambda \over h}p + p \\
2\Gamma\omega^{3}-\Gamma(1+p+2{\lambda \over h})\omega  
\end{array}
\right].
\end{equation}
Here, the first row are the real components and the second
row are the imaginary ones.
This equation can be solved for $x$ and $y$,
\begin{eqnarray}
x&=&{h\omega^6+[\Gamma^2h-2\lambda-(2+p)h]\omega^4
\over {4h\lambda[\omega^4+(\Gamma^2-2)\omega^2+1} ]} \nonumber \\
& & \nonumber \\
& & \; \; \; + {[h(1+2p)+2\lambda(1+p)-\Gamma^2h p -2
  \Gamma^2\lambda]\omega^2-[2\lambda +h]p
\over {4h\lambda[\omega^4+(\Gamma^2-2)\omega^2+1} ]}
\label{eq:x0}
\end{eqnarray}
and
\begin{equation}
y = -{\Gamma\omega \over 4 \lambda}+{\Gamma\omega(p-1) \over 2 h
  [\omega^4+
(\Gamma^2-2)\omega^2+1]}.
\label{eq:y0}
\end{equation}
The solution simplifies drastically when $p=1$,
\begin{eqnarray}
x & = & {h\omega^2-2\lambda-h \over 4\lambda h} \nonumber \\
y & = & - {\Gamma \omega \over 4 \lambda}.
\label{eq:p1}
\end{eqnarray}

With these formula we can calculate the decay length.  We use the fact that
$\cos(k+i\xi)=\cos k \cosh \xi - i \sin k \sinh \xi$.  Then,
\begin{eqnarray}
x  = & \Re\left\{\sin^2(z/2)\right\}= &{1 \over 2}(1-\cos k \cosh \xi)
\nonumber \\
y = & \Im\left\{\sin^2(z/2)\right\}= &{1 \over 2}\sin k \sinh \xi.
\label{eq:xy}
\end{eqnarray}

We can now solve for the $k$ and $\xi$.
The solution is not simple in the general case but when $y=0$, it simplifies
considerably.  When $k$ is zero,
\begin{equation}
\xi = \cosh^{-1}|1-2x|,
\label{eq:xiy0}
\end{equation} 
and the solutions decay exponentially.
When $\xi=0$ we recover the normal modes of the system and the
frequencies are given by the dispersion relation Eq.~(\ref{eq:res}). 

When $y \neq 0$ the solutions 
decay exponentially for all frequencies. This is due to the damping and 
is crucial for the existence of the resonant breather and chaotic
breather solutions found in Sec.~\ref{sec:tB} and Sec.~\ref{sec:tA}.

Fortunately, our ladders are underdamped so the $\Gamma \rightarrow 0$ limit
is appropriate.  Then, from Eq.~(\ref{eq:x0}) and Eq.~(\ref{eq:xiy0}),
\begin{equation}
\xi = \cosh^{-1} \left| {-h\omega^4 +(2\lambda h+ h(p+1)+2\lambda)\omega^2-2\lambda h-2 \lambda p -p h
\over 2\lambda h(\omega^2 -1)} \right|
\label{eq:decayg0}
\end{equation}
As $p \rightarrow 1$, this simplifies to
\begin{equation}
\xi = \cosh^{-1}\left|{2\lambda(h+1)+h \over 2 \lambda h} -{\omega^2 \over 2 \lambda}\right|.
\end{equation}
In the opposite limit of $\Gamma \rightarrow \infty$, $1/\xi$ approaches $0$.

Eq. (\ref{eq:decayg0}) gives the result for the decay
length of linearized excitations in the ladder when $\Gamma=0$.
Another limit where the equations simplify occurs
when $\omega \rightarrow 0$.

In the limit of $\omega \rightarrow 0$ we get
\begin{equation}
\xi = \cosh^{-1}\left({2\lambda h+p(2\lambda+h) \over 2 \lambda h} \right),
\label{eq:ozlim}
\end{equation}
and it can be verified that this result is
independent of $\Gamma$.

We have taken the $\omega \rightarrow 0$ and the $\Gamma \rightarrow 0$
limits.  There is one other important limit when inductances can be
neglected ($\lambda \rightarrow \infty$).  Then, with $\lambda \rightarrow
\infty$ and $p \rightarrow 1$, the decay length is 
\begin{equation}
\xi = \cosh^{-1}\left({h+1 \over h}\right)
\end{equation}
and it can also be verified that this is independent of $\Gamma$.
Also as $\lambda \rightarrow 0$, $1/\xi$ approaches $0$.

The decay length given by Eq. (\ref{eq:ozlim}) describes the
decay of the DC flux in the array.  For instance,
using the parameters in Fig.~\ref{fig:flux} we calculate
that $1/\xi=0.32$ and this agrees with the simulations.  This results
also implies that the decay is exponential and
that this exponential localization has an upper bound.  The
decay length $1/\xi$ is always less than  $1/\cosh^{-1}\{(h+p)/h\}$
if $\lambda \gg h$ or $1/\cosh^{-1}\{(2\lambda+p)/2\lambda\}$
if $h \gg \lambda$.

\section{Numerical and analytical study of single-site DB solutions}

\subsection{Regions of existence of single-site DB solutions}
\label{sec:nas}

In section~\ref{sec:brsims} we numerically simulated the behavior of
the DB solutions found in the experiments.  Such experiments were done
at moderate to small values of the damping, anisotropy and penetration
depth ($\Gamma \simeq 0.1$, $h \simeq 0.25$ and $\lambda \simeq
0.05$). It is also important to study the existence and behavior of
the localized solutions at other values of the parameters and to
estimate the critical values at which DB solutions
destabilize. Varying the temperature we studied experimentally the
behavior of the solutions at different values of the damping
$\Gamma$. The results were presented in Fig.~\ref{fig:temp}, where $h
\simeq 0.25$ and $\lambda \simeq 0.05$. For these values of the
parameters good agreement was found with the theoretical predictions.

The equations found in section~\ref{sec:cir} allow for a calculation
of the IV curves and the maximum and minimum values of external
currents supporting DB. For $R_b \gg R_h$ and single-site breather
solutions we find
\begin{equation}
\frac{I_a}{N} = \left (1+\frac{2h}{s} \right )\frac{V_v}{R_v}
\label{eq:pred1}
\end{equation}
for the IV curves and
\begin{eqnarray}
\frac{I_-}{N I_{cv}} & = & \left (2h+s \right )\frac{4}{\pi}\Gamma, \nonumber \\
\frac{I_+}{N I_{cv}} & = & \frac{2h+s}{3h+s}
\label{eq:pred2}
\end{eqnarray}
for the currents.

\begin{figure}[t]
\epsfxsize=4.5in
\centering{\mbox{\epsfbox{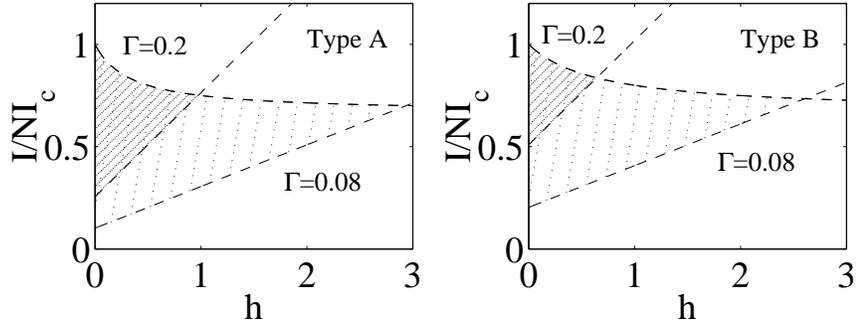}}}
%\vspace{0.15in}
\caption[Existence of breather equation]{Prediction of
Eqs.~(\ref{eq:pred2}) for $I_{+}$ and $I_{-}$ as a function of $h$ for
single-site type A (left) and type B (right) solutions and two values
of the damping. Lightly hatched region corresponds to $\Gamma=0.08$
and the densely one to $\Gamma=0.2$}
\label{fig:diag_eq1}
\end{figure}

\begin{figure}[t]
\epsfxsize=4.5in
\centering{\mbox{\epsfbox{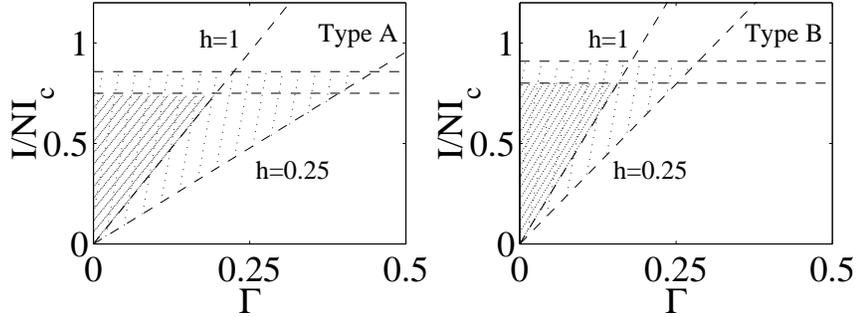}}}
%\vspace{0.15in}
\caption[Existence of breather equation]{Prediction of
Eqs.~(\ref{eq:pred2}) for $I_{+}$ and $I_{-}$ as a function of
$\Gamma$ for single-site type A (left) and type B (right) solutions
and two values of the anisotropy. Lightly hatched region corresponds to
$h=0.25$ and the densely one to $h=1.0$)}
\label{fig:diag_eq2}
\end{figure}

Figs.~\ref{fig:diag_eq1} and~\ref{fig:diag_eq2} show the predictions
given by the circuit model. The size of the existence regions decrease
rapidly when the damping or the anisotropy increase. 
On the other hand, if the damping is small enough
we should find localized
solutions at large values of $h$.
Also, the existence region is larger for
type A solutions.

This simple model, however, does not account for any dependence of the curves
with the parameter $\lambda$. This is an important limitation of the model and
we have been unable to develop a more complete, yet still simplified, approach
which incorporates $\lambda$.  We have confirmed in the numerical simulations
that $\lambda$ affects our predictions in two important ways. First, it
affects the value of the array retrapping current.  The value used in our
circuit models has been calculated from a single junction and should be
corrected by $\lambda$ in the case of the array. Indeed, some of the curves we
will show below can be fitted assuming a simple linear dependence of the
retrapping current with $\lambda$.  Second, as studied in the previous
section, it governs the values of the voltage at which resonances between the
breather and the normal modes of the array play an important role. Roughly
speaking, the resonances split the diagrams in two different regions: The
small and the large $\lambda$ regions.  When $\lambda$ is small, the resonance
frequency is smaller than the DB frequency, and when $\lambda$ is large the
resonance frequency is larger.  Thereby, complications of damped resonances
between the DB and the lattice eigenmodes are avoided in these limits.

Far from the resonance values the effect of $\lambda$ is a small
correction to our IV curves. This is shown by the numerical
simulations. See for instance Figs.~\ref{fig:pltivB}
and~\ref{fig:pltivA}, where IV curves numerically integrated agree
quite well with the predictions of the model, Eq. (\ref{eq:pred1}).

\begin{figure}[tb]
\epsfxsize=3.0in
\centering{\mbox{\epsfbox{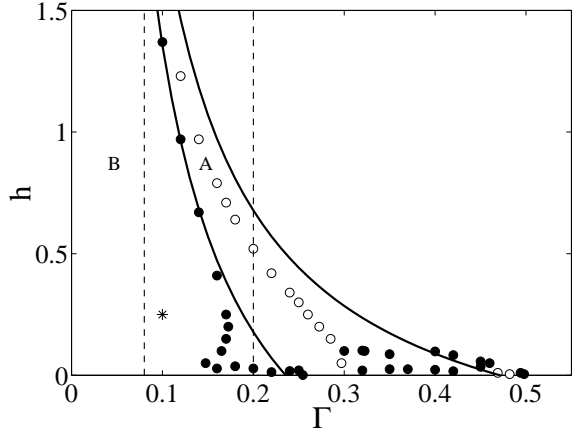}}}
%\vspace{0.15in}
\caption{Numerical calculation of the existence region of single-site DB
when $\lambda=0.04$ and $I=0.6$. Open circles correspond to type-A and solid
circles to type-B solutions. Vertical lines correspond to cuts show in
Fig.~\ref{fig:diag_hl} and the asterisk to the experiments.}
\label{fig:diag_hg}
\end{figure}

\begin{figure}[tb]
  \epsfxsize=3.0in \centering{\mbox{\epsfbox{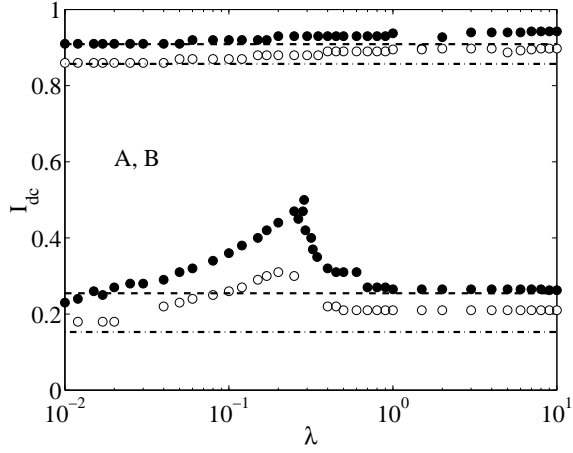}}}
%\vspace{0.15in}
\caption{Numerical calculation of the existence region of single-site DB 
  when $\Gamma=0.08$ and $h=0.25$. Open circles correspond to type-A and solid
  circles to type-B solutions. Horizontal lines correspond to the predictions
  of the circuit model and the asterisk to the experiments.}
\label{fig:diag_il}
\end{figure}

\begin{figure}[tb]
\epsfxsize=4.5in
\centering{\mbox{\epsfbox{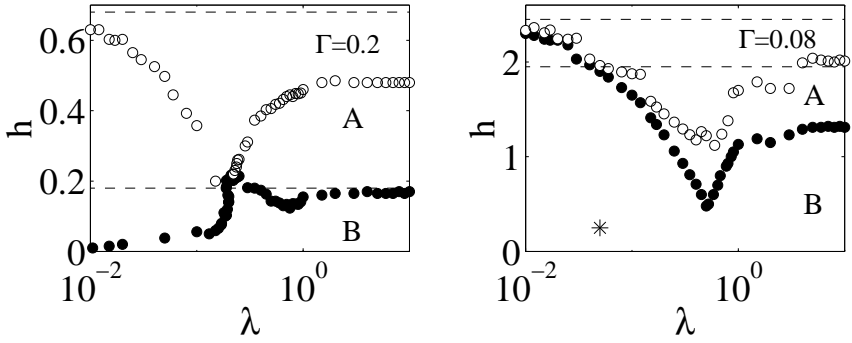}}}
%\vspace{0.15in}
\caption{Numerical calculation of the existence region of single-site DB when
  $I=0.6$ and $\Gamma=0.2$ (left) and $\Gamma=0.08$ (right). Open circles
  correspond to type-A and solid circles to type-B solutions. Horizontal lines
  correspond to the predictions of the circuit model and the asterisk to the
  experiments.}
\label{fig:diag_hl}
\end{figure}

We have done numerical simulations based on Eqs. (\ref{eq:ladder}) with $f=0$
in order to study the $\lambda$ dependence of the breather existence region.
The results are presented in Figs.~\ref{fig:diag_hg}
through~\ref{fig:diag_hl}.  In these diagrams we show the maximum and minimum
values of the parameters for which a localized solution has been numerically
found. As we will see, in some cases the characterization of the solutions
inside the existence regions is quite complex in which several resonances and
transitions between periodic and aperiodic states appear. In the figures we
have also marked the values of the parameters at which the experiments were
done, all far from these {\em problematic} areas and belonging to a region of
the diagram where both type A and type B breathers are predicted to exist.

The data were calculated by integrating the governing equations
with a small quantity of noise.
We start with a type B rotobreather and $h \sim 0.001$. As we increase $h$,
type B solutions
become unstable and the solution evolves to a type A rotobreather. As we further
increase $h$ this rotobreather becomes unstable and the system usually jumps to
either a superconducting or a whirling state. 
To verify that our method is accurate, we have
calculated Floquet multipliers for periodic rotobreather solutions and found
results consistent with those shown.

Fig.~\ref{fig:diag_hg} shows the existence regions 
in the anisotropy versus damping
plane when $\lambda=0.04$ and $I=0.6$.  When $h$ is 
large the agreement with the predictions of Eq. (\ref{eq:pred2}) is good.
There is an abrupt deviation of the curve for type B solutions at
small values of $h$ and a region at larger $\Gamma$ where new type-B solutions
appear. Vertical lines correspond $\Gamma=0.08$ and $\Gamma=0.2$, studied in
Fig.~\ref{fig:diag_hl}.  The  asterisk corresponds to the experimental
parameters.

Fig.~\ref{fig:diag_il} shows the existence regions
in the current versus $\lambda$ plane
when $\Gamma=0.08$ and $h=0.25$. These are the values of $h$ and $\Gamma$ in
our experiments.  We can see that Eq. (\ref{eq:pred2}) gives a good
estimation of the maximum and minimum values of the current for localized
solutions, except for the case of the
minimum current with a moderate $\lambda$.  In this case,
resonances and other dynamical effects
cause a substantial deviation from our simple model.

Fig.~\ref{fig:diag_hl} shows the diagram in the anisotropy versus
$\lambda$ plane for $I=0.6$ and $\Gamma=0.2$ (left) and $0.08$
(right). As expected from Fig.~\ref{fig:diag_eq1} localized solutions
exist at larger values of $h$ when $\Gamma$ is smaller. Unexpectedly,
at $\Gamma=0.2$ and small values of $\lambda$ type-B solutions exist
only for small values of $h$. This behavior is also described in
Fig.~\ref{fig:diag_hg}. The asterisk in the $\Gamma=0.08$ figure
approximately correspond to the value of the parameter where our
experiments were done.  

Clearly, the existence diagrams are
complex and further research is necessary to fully understand them.
However, these diagrams show that DB solutions might be understood
within our simple model at limiting values of small and large 
$\lambda$.  We will use these limits in the numerical 
and analytical analysis of type B and type A
solutions in the following sections.

\subsection{Type B breathers}
\label{sec:tB}

\subsubsection{Simulations}

\begin{figure}[tb]
\epsfxsize=4.5in
\centering{\mbox{\epsfbox{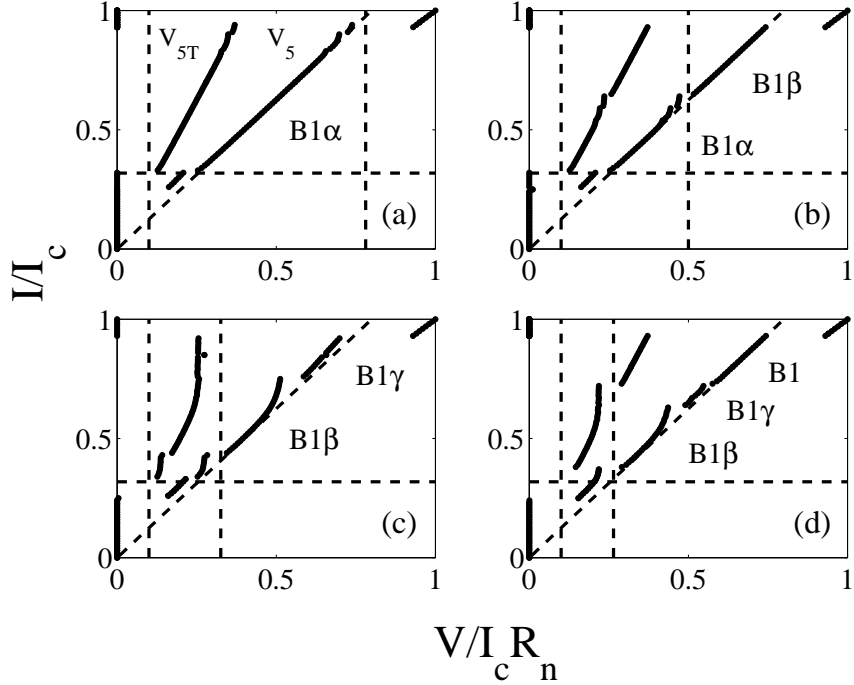}}}
%\vspace{0.1in}
\caption[Simulated ladder IV's for type B breather vs $\lambda$]{
Simulated IV's for $9\times1$ ladder of type B breather as a function
of $\lambda$. $h=0.25$, $\Gamma=0.1$, $f=0$ and (a) $\lambda=5$, (b)
$\lambda=2$, (c) $\lambda=0.8$, and (d) $\lambda=0.5$. The labels
indicate different type B breather solutions.  The vertical dashed
lines are $\omega_{+}(\lambda)$ and $\omega_{-}=1$ from
Eq. (\ref{eq:res}).  The horizontal dashed line is $I_{-}$ from
Eq. (\ref{eq:minI2}) when  $R_b \gg R_h$ and the diagonal dashed
line is from Eq. (\ref{eq:ciriv}) when $R_b \gg R_h$. }
\label{fig:pltivB}
\end{figure}

Fig.~\ref{fig:pltivB} shows typical simulated IV's at different $\lambda$'s
for single-site type B breathers.  The horizontal dashed line is the minimum
retrapping current expected from Eq. (\ref{eq:minI2}) when $R_b \gg R_h$ in
the circuit model.  The vertical dashed lines are the two resonant voltages of
the ladder as approximated with Eq. (\ref{eq:res}) when $z=\pi$ (the largest
lattice frequency). As can be seen in the IV's, $\omega_{+}$ gives a good
approximation of the location of the resonant structure.  Also, $\omega_{+}$
is always larger than $\omega_{-}$.  The diagonal dashed line is the expected
voltage of the fifth vertical junction calculated from the DC circuit model,
Eq. (\ref{eq:ciriv}) with $R_b \gg R_h $.  The voltages of the fifth vertical
junction and the fifth top horizontal junction are plotted and are usually
related by $V_{5}=2V_{5T}$.  Here, we have used the simple RCSJ model without
a subgap resistance and we have set $\Gamma=0.1$, the experimental value
measured from the subgap resistance.

\begin{figure}[tb]
\epsfxsize=5in
\centering{\mbox{\epsfbox{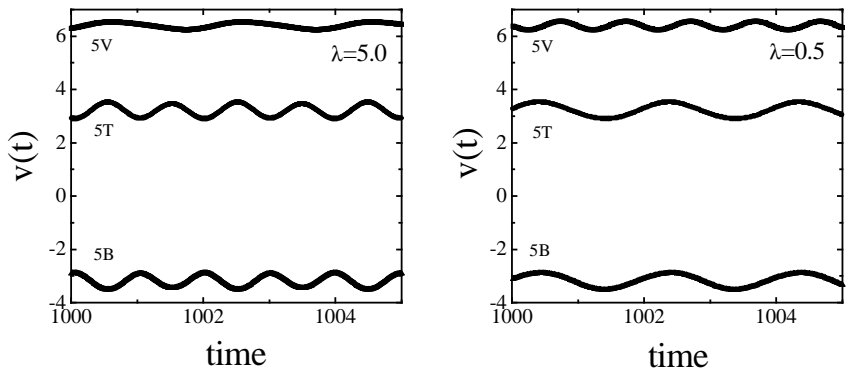}}}
%\vspace{0.1in}
\caption[vt]{Time evolution of the time derivative of the phase,
$v(t)=d \varphi /dt$, for the large $\lambda$ type B solution labeled
B1$\alpha$ in Fig.~\ref{fig:pltivB}(a) and the small $\lambda$ type B
solution depicted as B1 in Fig.~\ref{fig:pltivB}(d).  We plot $v(t)$
for the rotating vertical junction (5V) and two neighbor horizontal
junctions (5T and 5B). Here, $h=0.25$, $\Gamma =0.1$ and $I=0.8$.}
\label{fig:vt}
\end{figure}

Fig.~\ref{fig:pltivB}(a) shows an IV when $\lambda=5$. It corresponds to the
large $\lambda$ regime. At these parameter values the upper resonance is above
any of the junction voltages.  However, we still see some resonant behavior at
$I \approx 0.9$.  The solution when $I=0.8$ is shown in
Fig.~\ref{fig:vt}(left).  This type B breather is fully up-down symmetric in
that $\varphi_j^t(t)=-\varphi_j^b(t)$. Also, the averaged voltage of the
horizontal junction is always half of the voltage of the vertical junction.
We will label this type of solution as B1$\alpha$.  As we decrease the
current, the breather becomes unstable at $I=0.32$ as predicted by the
retrapping model, Eq. (\ref{eq:minI2}), and a type A breather forms which
itself becomes unstable at $I=0.26$.

Fig.~\ref{fig:pltivB}(b) shows an IV when $\lambda=2$.  We see that the $V_5$
branch is now separated into two parts by the $\omega_{+}$ resonance while
$V_{5T}$ is still below the resonance.  The solution when all the voltages are
below the resonance is still B1$\alpha$.  We will label as B1$\beta$ the
breather solution where $V_{5T}$ is below and $V_{5}$ is above the resonance.
There is also a hysteresis loop that forms at the resonance that is not shown
in the figure.  We find that, as the B1$\alpha$, the B1$\beta$ solution is
also up-down symmetric, but in the case of the B1$\beta$ there exists a phase
difference in between the velocity of the vertical and the horizontal
junctions.

When $\lambda=0.8$ we see the IV shown in Fig.~\ref{fig:pltivB}(c).
We find that we can interpret the $\beta$ solution as a resonant type
B breather.  Below the resonance, there is a small remnant of the
$\alpha$ breather.  When the $\beta$ solution becomes unstable but
$V_{5T}$ is still below the resonance, there is another type B
breathers which we have labeled B1$\gamma$.  An unusual signature of
this breather is that $V_{5}\neq 2V_{5T}$.  Further analysis shows
that the $\gamma$ solution is an aperiodic type B breather.
There is also a hysteresis loop at the resonance that, depending on
the parameters, might surround the aperiodic breather.  This hysteresis
would make it difficult to experimentally access this attractor
by only using the applied current.

We can continue to decrease $\lambda$.  Fig.~\ref{fig:pltivB}(d) shows
an IV when $\lambda=0.5$.  The upper resonance has now divided the
voltage of $V_{5T}$ into two branches.  Below the resonance we get the
solutions described above.  We see that the $\beta$ solution is
getting ``squeezed'' by the $\omega_{+}$ resonance and the retrapping
current.  Indeed, there is a critical value of $\lambda$ where the
type $\beta$ solution disappears.

Above the resonance, we find a solution labeled B1.  This small
$\lambda$ regime is the same as in the experiments.  The single-site
breathers measured in our ladders are of this B1 type while the other
breathers described above were only found numerically.  In
Fig.~\ref{fig:vt}(right) we see that this B1 solution is not up-down
symmetric although $V_{5T}=V_5/2=-V_{5B}$.

\begin{figure}[tb]
\epsfxsize=\figsize
\centering{\mbox{\epsfbox{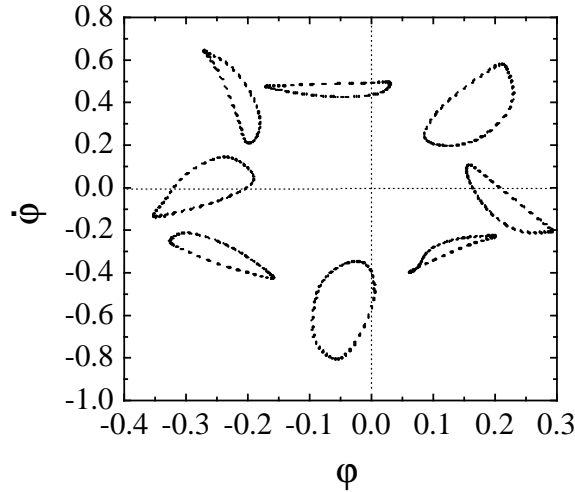}}}
%\vspace{0.15in}
\caption{Poincar\'e sections of the third top junction (3T) for a
B1$\gamma$ solution. The phases are shown at times $t=t_0+n\tau$ where
$\tau=2\pi/V_5$ and $h=0.25$, $\Gamma=0.1$, $\lambda=0.5$, and $I=0.7$.}
\label{fig:quasi}
\end{figure}

We have done a Poincar\'e section analysis of the aperiodic solution
B1$\gamma$ at the same parameter values as Fig.~\ref{fig:pltivB}(d) In
Fig.~\ref{fig:quasi} we show the value of $\dot{\varphi}$ versus
$\varphi$ at times equal to $t_0+n\tau$ where $\tau=2\pi/V_5$. The
simplest periodic type B solutions 
have a period $T=2\tau$ since horizontal junction is
half the voltage of the vertical junction. The solution shown
in Fig.~\ref{fig:quasi} seems to be quasiperiodic with two incommensurate
frequencies, one of which seems to have a period equal to
$4T$. 

There are many other type B breathers that were found numerically at
other values of the parameters but are not discussed here.  For instance, 
there is a family of solutions which is not left-right symmetric.  However,
we shall focus our discussion to left-right symmetric solutions that
have the above characteristics.

\begin{figure}[tb]
\epsfxsize=5in
\centering{\mbox{\epsfbox{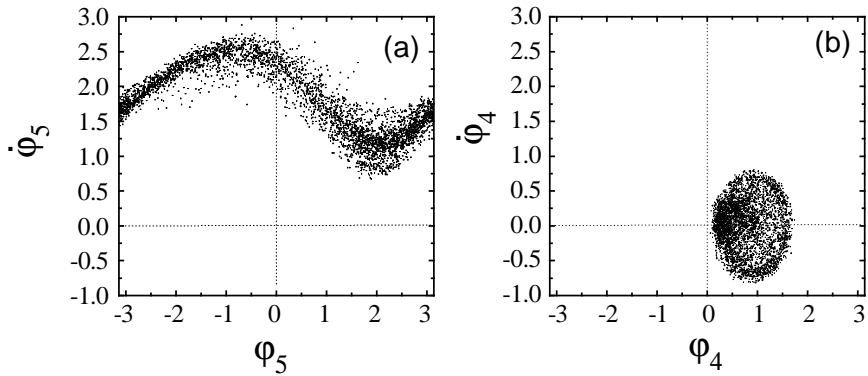}}}
%\vspace{0.15in}
\caption{Poincar\'e sections of the fourth and fifth 
vertical junctions for a type B
chaotic solution. The phases are shown at times $t=t_0+n\tau$ where
$\tau=2\pi/V_5$ and $h=0.15$, $\Gamma=0.2$ and $\lambda=0.2$ and
$I=0.6$. (a) Shows the sections for the rotating vertical junction 5
and (b) for its first neighbor 4.}
\label{fig:chaotic}
\end{figure}

At other values of the parameters, the IV curves show chaotic localized
solutions.  Fig.~\ref{fig:chaotic} shows Poincar\'e section for vertical
phases in the case of a chaotic localized type B solution. The values of the
parameters for this solution are $h=0.15$, $\lambda=0.2$, $\Gamma=0.2$ and
$I=0.6$.  Such chaotic region can be located in the central part of
Fig.~\ref{fig:diag_hl}.

\subsubsection{Analysis}
\label{Banal}

In this section, we will use a harmonic
balance technique to derive some analytical descriptions of our DB.
When studying periodic solutions it is almost always easier to
work in Fourier space. 
Harmonic balance is a technique where the variables are decomposed
into their Fourier components and substituted back into the governing
equations.  This creates a large set of coupled algebraic equations.
If the governing equations are linear then each resulting algebraic
equation is independent and the full system is easily solved.
However, nonlinearities tend to mix harmonic components.  If the
mixing effect is large then the resulting set of algebraic equations
is usually  intractable.  However, in our underdamped Josephson arrays
the capacitances act as filters that allow the transmission of only
a few frequencies.  Typically it is only one frequency.  In
this case, we can truncate the set of algebraic equations and
a harmonic balance technique can provide useful
approximations.

From simulations almost all breather solutions have left-right
symmetry (and this is always true for average voltages).  We can then make
a transformation from the full ladder to what we call a half-ladder.
A half-ladder is ladder where the breather is now on the first
junction and we assume left-right symmetry.

To make a mapping of the equations from a half-ladder to a
full ladder we first consider placing a breather in junction 5 of a
9-junction ladder.  Current conservation at that node yields,
\begin{equation}
I^v_5=I^t_5-I^t_4+I^e.
\end{equation}
Mirror symmetry implies that $I^t_5=-I^t_4$ thereby
current conservation becomes
\begin{equation}
I^v_5=2I^t_5 + I^e
\end{equation}
If we have a half ladder, current conservation
at the first node (labeled 5 for comparison) yields
\begin{equation}
I^v_5=I^t_5 +I^e
\end{equation}
To make
a mapping between the circuit equations, 
we need to multiply $I^t_5$ by 2.  Since
$I^t=h{\cal N}(\varphi)$, this is
equivalent to setting $h_{\rm half\,ladder}=2h_{\rm ladder}$.
However, we want to maintain the same flux pattern in
the half-ladder so
fluxoid quantization must remain unchanged.  This is
simply done by setting
$\lambda_{\rm half\,ladder}=2\lambda_{\rm ladder}$.

Since the breather solution is highly localized we can,
as a first approximation, neglect every cell except the first
one.  The resulting reduced system of equations can be written
as
\begin{eqnarray}
{\cal N}(\varphi^t)&=& - {\cal N}(\varphi^b)      
\label{eq:cellTopDown} \\
h_c{\cal N}(\varphi^t)+{\cal N}(\varphi^l)&=&i_{\rm ext}    
\label{eq:cellKCLl} \\
-h_c{\cal N}(\varphi^t)+{\cal N}(\varphi^r)&=&i_{\rm ext}   
\label{eq:cellKCLr} \\
-\varphi^t+\varphi^b+\varphi^l-\varphi^r&=&{h_c\over \lambda_c}
{\cal N}(\varphi^t)
\label{eq:cellFQ}
\end{eqnarray}
where $h_c=2h_{\rm ladder}$ and $\lambda_c=2\lambda_{\rm ladder}$.

Most type B breathers have two voltages in the system corresponding to
two frequencies.  Our simulations indicate that the voltage of the
horizontal rotating junction can be approximately decomposed as
\begin{equation}
v^{\{t,b\}} = \pm {\omega \over 2} + 
a^{\{t,b\}}\cos \left ({\omega \over 2}t+\theta^{\{t,b\}}_a \right) +
b^{\{t,b\}}\cos \left({\omega }t+\theta^{\{t,b\}}_b \right)
\end{equation}
Similarly for the rotating vertical junction,
\begin{equation}
v^l = {\omega }  + 
a^l\cos \left ({\omega \over 2}t+\theta^l_a \right)  
 +  b^l\cos \left({\omega }t+\theta^l_b \right).
\end{equation}
For the librating junction
\begin{equation}
v^r = 
a^r\cos\left({\omega \over 2}t+\theta^r_a\right) 
+b^r\cos\left({\omega}t+\theta^r_b\right).
\end{equation}

We integrate to get the phases and
use exponential notation.
Our phases are then
\begin{eqnarray}
\varphi^{t} & = &
\Re \left \{ {\omega \over 2}t+k^{t}+{\pi \over 2}
-i A^{t} e^{i{\omega \over 2}t} 
-i B^{t} e^{i{\omega}t} \right \} \nonumber  \\
\varphi^{b} & = &
\Re \left \{ -{\omega \over 2}t+k^{b}+{\pi \over 2}
-i A^{b} e^{i{\omega \over 2}t} 
-i B^{b} e^{i{\omega}t} \right \}  \nonumber  \\
\varphi^l & = &
\Re \left \{ {\omega }t+k^l+{\pi \over 2}
-i A^l e^{i{\omega \over 2} t}
-i B^l e^{i{\omega }t} \right \} \nonumber  \\
\varphi^r & = & \Re \left \{  k^r
-i A^r e^{i{\omega \over 2} t}
-i B^r e^{i\omega t} \right \}
\end{eqnarray}
where $A^x=(2a^x/ \omega) e^{i\theta_a^x}$ and $B^x=(b^x/ \omega)
e^{i\theta_b^x}$ with $x=t, b, l$ and $r$.
By convention we will take the real parts for the actual
variables.  Also, the integration constant has been taken to be
$k^{\{t,b,l\}}+ \pi/2$ for the rotating junctions.

To substitute
into our governing equations we must 
linearize the sine term.  Substituting our
phase ansatz into the sine term
would yield an infinite Fourier
series whose coefficients are Bessel 
functions of the amplitudes.  In this sense
the sine terms mix all of the harmonics.  In the
present case, our amplitudes are small and
we can linearize
$\cos x =1$ and $\sin x =x$.
The sine for the
top rotating phase
is approximated as
\begin{equation}
\sin \varphi^t =  \Re \{
e^{i({\omega \over 2} t + k^t)}-{A^t \over 2}e^{-ik^t} 
+ {A^t \over 2}e^{i(\omega t + k^t)}
- {B^t \over 2}e^{i({\omega \over 2} t - k^t)} \}
\end{equation}
and for the bottom phase
\begin{equation}
\sin \varphi^b =  \Re \{
e^{-i({\omega \over 2} t - k^b)}+{A^b \over 2}e^{ik^b}
- {A^b \over 2}e^{i(\omega t - k^b)}
+ {B^b \over 2}e^{i({\omega \over 2} t + k^b)} \}.
\end{equation}
The sine
of the vertical rotating phase is  approximated as
\begin{equation}
\sin \varphi^l =  \Re \{
e^{i({\omega } t + k^l)}
 -  {A^l \over 2}e^{-i({\omega t \over 2} + k^l)} 
 - {B^l \over 2}e^{-ik^l} \}.
\end{equation}
For the librating phase
\begin{equation}
\sin \varphi^r  =  \Re \{ -ie^{ik^r} 
 - i({A^r} e^{i{\omega \over 2} t }    
 + {B^r } e^{i{\omega} t } )\cos  k^r  \}.
\end{equation}
We have neglected all terms of frequencies not
equal to $\pm \omega$ or $\pm \omega/2$.

If we substitute our ansatz into our governing equations
and expand the sine terms as indicated above, we transform
the original 4 differential equations into a
linear system of 20 equations with
21 unknowns.  This extra degree of freedom
is associated with the time translational invariance
of the equations. 

In principle it is possible
to solve the full algebraic system by
fixing one of the unknowns,
but here we will 
just estimate the amplitude oscillations of the
breather.  
To make headway, we will use some trends found
in the simulations to reduce the number of unknowns and
take the limit $\Gamma=0$.

We first calculate $|A^t|$ and approximate
the phases from the numerical simulations.
From simulations, we find that as
$\lambda \rightarrow \infty$, the phase difference
of the $\omega/2$ harmonic between the waveforms
of $\ddot{\varphi}^t$ and $\sin \varphi^t$ is
$\pi$.  We also find that the phase 
difference of the $\omega/2$ harmonic
between the waveforms of $\varphi^t$ and $\varphi^l$ is zero and
between the waveforms of $\varphi^l$ and $\varphi^r$ is $\pi$.  These
phase relations can be used to reduce the number of unknowns
in the algebraic system.

One solution to Eq.~(\ref{eq:cellTopDown})
obeys the up-down symmetry: $B^t=-B^b$ and $A^t=-A^b$.
We can combine the $\omega/2$ harmonic parts of
Eq. (\ref{eq:cellKCLl}), Eq. (\ref{eq:cellKCLr}), and
Eq. (\ref{eq:cellFQ}) and solve for $|A^t|$.  This yields
\begin{equation}
|A^t|=\left | {2h_c\lambda_c-{h_c }({\omega / 2})^2 \over
({\omega / 2})^2 (2\lambda_c+2h_c\lambda_c-{h_c }({\omega / 2})^2)} \right |
\label{eq:At}
\end{equation}
Using the same technique to approximate phases
using simulations for the $\omega$ harmonics, we find
\begin{equation}
|B^r|=\left | {1 \over
{\omega^2 } (2+2/h_c-{\omega^2 }/\lambda_c)} \right |
\label{eq:Br}
\end{equation}
with $h|B^t|=|B^r|$.  We can then find all of the
amplitudes of the harmonics in terms of $\omega$.

As $\lambda$ decreases the breather enters a resonant regime.  In this
regime the phase relationships change from the above limit.  We now
find the phase difference of the $\omega/2$ harmonic of
the waveform of $\ddot{\varphi}^t$ and $\sin \varphi^t$ is zero, the phase difference
between the waveforms of $\varphi^t$ and $\varphi^l$ 
is $\pi$, and between the waveforms of $\varphi^l$ and
$\varphi^r$ is also $\pi$.  Substituting the new ansatz in the governing
equations results in the same equation as Eq.~(\ref{eq:At}) but with an
overall negative sign.  The same occurs with the $B^r$ amplitude.  In
the resonant regime Eq.~(\ref{eq:Br}) just changes sign.

In the $\lambda \rightarrow 0$, the equations simplify.  From
Eq. (\ref{eq:cellFQ}) we see that ${\cal N}(\varphi^t)$ must tend to zero.
This implies that
\begin{equation}
|A^t|={1 \over (\omega/2)^2}.
\label{eq:Atlsmall}
\end{equation}
This also puts strong constraints on $A^r$ and $A^l$.  One possible
solution is $A^r=0$ and $A^l=0$.  Then to satisfy Eq.~(\ref{eq:cellFQ}),
$A^t=A^b$ contrary to the expected up-down symmetry.
However, this solution is consistent with Eq. (\ref{eq:cellTopDown}) and
it is the stable solution we find numerically for small $\lambda$.

\begin{figure}[t]
\epsfxsize=4.0in
\centering{\mbox{\epsfbox{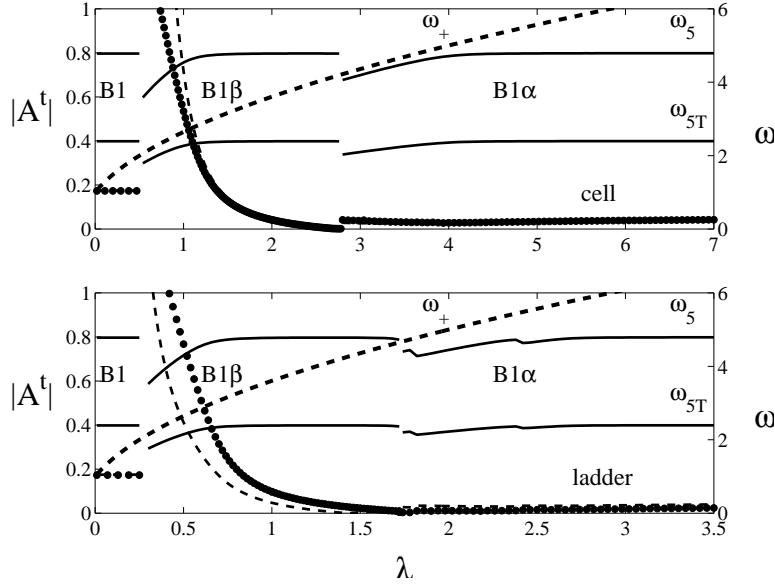}}}
%\vspace{0.15in}
\caption[Calculated type B breather amplitudes]{ Solid circles are
simulated $A^t$ while dashed line is calculated $A^t$ from
Eq. (\ref{eq:At}) for $\lambda$ large and Eq. (\ref{eq:Atlsmall}) for
$\lambda$ small.  Solid lines is $\omega$ found in the simulation.
  Here $\Gamma=0.1$ and
$h=0.25$.}
\label{fig:typeBamps}
\end{figure}

Fig.~\ref{fig:typeBamps} shows a summary of the analytical results and
how they compare to the simulations.  The top graph is a simulation
for a cell while the bottom graphs shows results for our ladder.
Both system have $\Gamma=0.1$.  For each system, we have excited a
type B breather at $i_{\rm ext}=0.5$ and $\lambda=10$.  We plot the
value of $|A^t|$ and $\omega$ as we decrease $\lambda$.

As expected, both graphs are very similar when $\lambda_{\rm
cell}=2\lambda_{\rm ladder}$.  The top solid line is from the
simulated frequency of the rotating vertical junction.  The bottom
solid line is from the simulated frequency of the horizontal junction
which is precisely half of the vertical frequency.  The dashed line
shows the $\omega_{+}$ resonance.  Because there are two frequencies,
the resonance divides the solution space into three regions that
result in the three different solution found in Fig.~\ref{fig:pltivB}
and studied analytically using harmonic balance.

The solid circles in Fig.~\ref{fig:typeBamps} are the simulated $A^t$
and the dashed lines are the analytical estimate.  For small $\lambda$
we have the B1 solution whose amplitude is given by
Eq. (\ref{eq:Atlsmall}) and the predicted amplitude lies on top of the
simulated values.  For the resonance regime of solution B1$\beta$ the
amplitude is given by Eq. (\ref{eq:At}), and for the large $\lambda$
regime the amplitude is also given by Eq. (\ref{eq:At}).

For the simulations of a single cell, the harmonic balance gives a
very good approximation.  This is not too surprising, since the
analysis was performed for a single cell.  However, the bottom of
Fig.~\ref{fig:typeBamps} shows that the analysis is also valid for the
full ladder as long as we transform $\lambda$ and $h$.

\subsection{Type A breathers}
\label{sec:tA}

\begin{figure}[tb]
\epsfxsize=4.5in
\centering{\mbox{\epsfbox{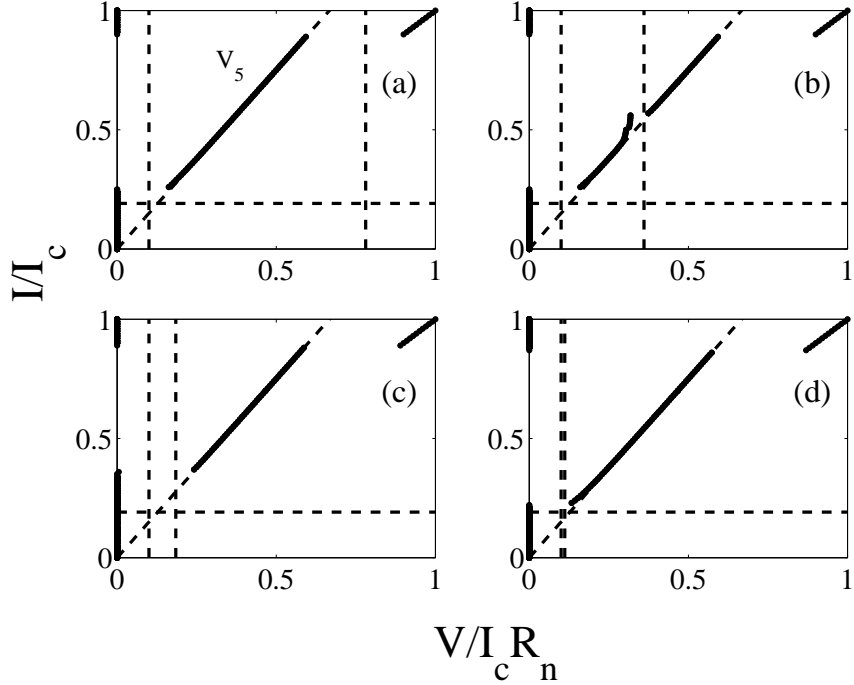}}}
%\vspace{0.1in}
\caption[Simulated ladder IV's for type A breather vs $\lambda$]{
Simulated IV's for $9\times1$ ladder of type A breather as a function
of $\lambda$.  Here, $\Gamma=0.1$, $f=0$ and (a) $\lambda=5$, 
(b) $\lambda=1$, (c) $\lambda=0.2$,
and (d) $\lambda=0.02$.  The vertical dashed lines are
$\omega_{+}(\lambda)$ and $\omega_{-}=1$ from Eq. (\ref{eq:res}).  The
horizontal dashed line is $I_{-}$ from Eq. (\ref{eq:minI2}) as
$R_b \gg R_h$ and the diagonal dashed line is from
Eq. (\ref{eq:ciriv}) also as $R_b \gg R_h$. }
\label{fig:pltivA}
\end{figure}

Fig.~\ref{fig:pltivA} shows typical simulated IV's as a function of $\lambda$
for type A breathers.  As in the previous discussion, the horizontal dashed
line is the expected retrapping current while the vertical dashed lines are
the two resonant voltages from Eq. (\ref{eq:res}).  Here the voltage of the
fifth vertical junction and the fifth horizontal junction are the same and
this makes the analysis of the solutions much simpler.  There are basically
just two solutions.

A harmonic balance technique can be used to calculate the 
amplitudes of the oscillating parts of the breather.
We use a similar approach to Sec.~\ref{Banal} but
approximate the phases with one frequency instead of two.
Our phases are then
\begin{eqnarray}
\varphi^{t} & = &
\Re \left \{ {\omega }t+k^{t}+{\pi \over 2}
-i A^{t} e^{i{\omega t}} \right \}\nonumber  \\ 
\varphi^{b} & = &
\Re \left \{ k^{b}
-i A^{b} e^{i{\omega }t} \right \}  \nonumber  \\
\varphi^l & = &
\Re \left \{ {\omega }t+k^l+{\pi \over 2}
-i A^l e^{i{\omega } t} \right \} \nonumber  \\
\varphi^r & = & \Re \left \{  k^r
-i A^r e^{i{\omega } t} \right \}.
\end{eqnarray}

When $\omega_{+}$ is above the junction voltages, as in
Fig.~\ref{fig:pltivA}(a) the solution is up-down symmetric with respect to the
amplitudes: $A^t=-A^b$, and $A^r\approx A^t$.  This means that all of the core
junctions of the breather have a small oscillating amplitude.  As $\lambda$
decreases the breather can resonates with the lattice eigenmodes as shown in
Fig.~\ref{fig:pltivA}(b).  When $\lambda$ is so small that $\omega_{+}$ is
below the breather frequencies we find a solution with $A^b\approx 0$ and
$A^r\approx 0$.

The analysis of the amplitudes is straight forward and
similar to what is done in Sec.~\ref{Banal}.  We
substitute the ansatz in the governing equations and
keep only DC and $\omega$ harmonics.   We then
use simulations to approximate some of the phase relations and
thereby reduce the number of equations. 

In the
$\lambda=0$ limit, we have from Eq.~(\ref{eq:cellFQ}) that ${\cal
N}(\varphi^t)=0$.  If we set $\Gamma=0$, this implies that
\begin{equation}
|A^t|={1 \over \omega^2}
\label{eq:AAtlsmall}
\end{equation}
In the opposite limit, we can again use Eq. (\ref{eq:cellFQ}).
We substitute the ansatz found in the simulations of
$A^t=-A^b$ and approximate $A^r$ as $A^t$, then
\begin{equation}
|A^t|=\left | {h/\lambda \over 3-h\omega^2/\lambda} \right |
\label{eq:AAt}
\end{equation}

\begin{figure}[t]
\epsfxsize=4.0in
\centering{\mbox{\epsfbox{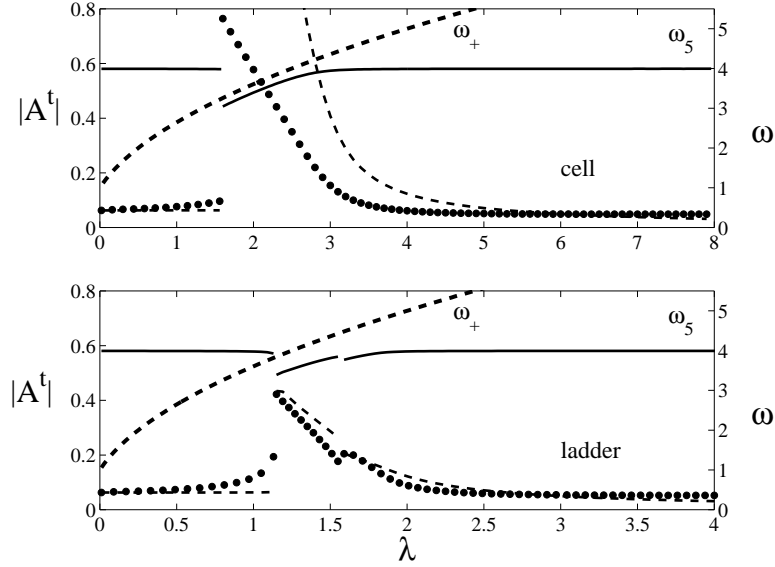}}}
%\vspace{0.15in}
\caption[Calculated type A breather amplitudes]{ Solid circles are
simulated $A^t$ while dashed line is calculated $A^t$ from
Eq. (\ref{eq:AAt}) for $\lambda$ large and Eq. (\ref{eq:AAtlsmall}) for
$\lambda$ small.  Solid line $\omega$ found in the simulation. 
Here $\Gamma=0.1$ and
$h=0.25$.}
\label{fig:typeAamps}
\end{figure}

Fig.~\ref{fig:typeAamps} shows a comparison the analytical results and
how they compare to the simulations.  The top graph is a simulation
for a cell while the bottom graphs shows results for the ladder.
Both system have $\Gamma=0.1$.  For each system, we have excited a
type A breather at $i_{\rm ext}=0.4$ and $\lambda=10$.  We plot the
value of $|A^t|$ and $\omega$ as we decrease $\lambda$.

As expected, both graphs are very similar when $\lambda_{\rm
cell}=2\lambda_{\rm ladder}$.  The solid line is from the simulated
frequency of the rotating vertical junction.  The dashed line shows
the $\omega_{+}$ resonance and we see that there are two types of
solutions: Breathers with frequencies above $\omega_{+}$, and
breathers with frequencies below $\omega_{+}$.

The solid circles in Fig.~\ref{fig:typeAamps} are the simulated $A^t$
and the dashed lines are the analytical estimate.  For small $\lambda$
the amplitude is given by Eq. (\ref{eq:AAtlsmall}) and the predicted
amplitude match the simulated values.  For the large $\lambda$ regime
the amplitude is given by Eq. (\ref{eq:AAt}) and the approximation
breaks down at the resonance.  In any case, we see that comparing a
breather in one cell to the breather in a ladder works quite well.

\subsection{Vortex patterns in breathers}
\label{sec:vor}

The DB that we observe in the ladder can also be understood as bound vortex
states. In Eq. (\ref{eq:vorticity}) we define the vorticity $n_j$ in a cell
as
\begin{equation}
n_j=
{1\over 2\pi}\{[\varphi^v_j]-[\varphi^v_{j+1}]-[\varphi^t_j]+[\varphi^b_j]\}
+f^{\rm ind}_j
\label{eq:vorticity2}
\end{equation}
where $[\varphi]$ represents the phases modulus $2\pi$, and
$f^{\rm ind}_j=i^m_j/2\pi\lambda$.
As we will show, $n_j$ changes in time differently for the three
analyzed solutions. However, we want to stress that the magnetic field
flux, which is due to the mesh currents in the ladder, librate around a mean
value as 
shown in Fig.~\ref{fig:flux} and do not circulate. Only the fluxoid, or
vorticity, which is the quantized quantity, moves from cell to cell.
This behavior is different for the cases of type B or type A
solutions and in the case of type B solutions for small or large values of
$\lambda$.

\subsubsection{Large $\lambda$ type B solutions}
\label{sec:vorBlarge}

\begin{figure}[tb]
\epsfxsize=3.2in
\centering{\mbox{\epsfbox{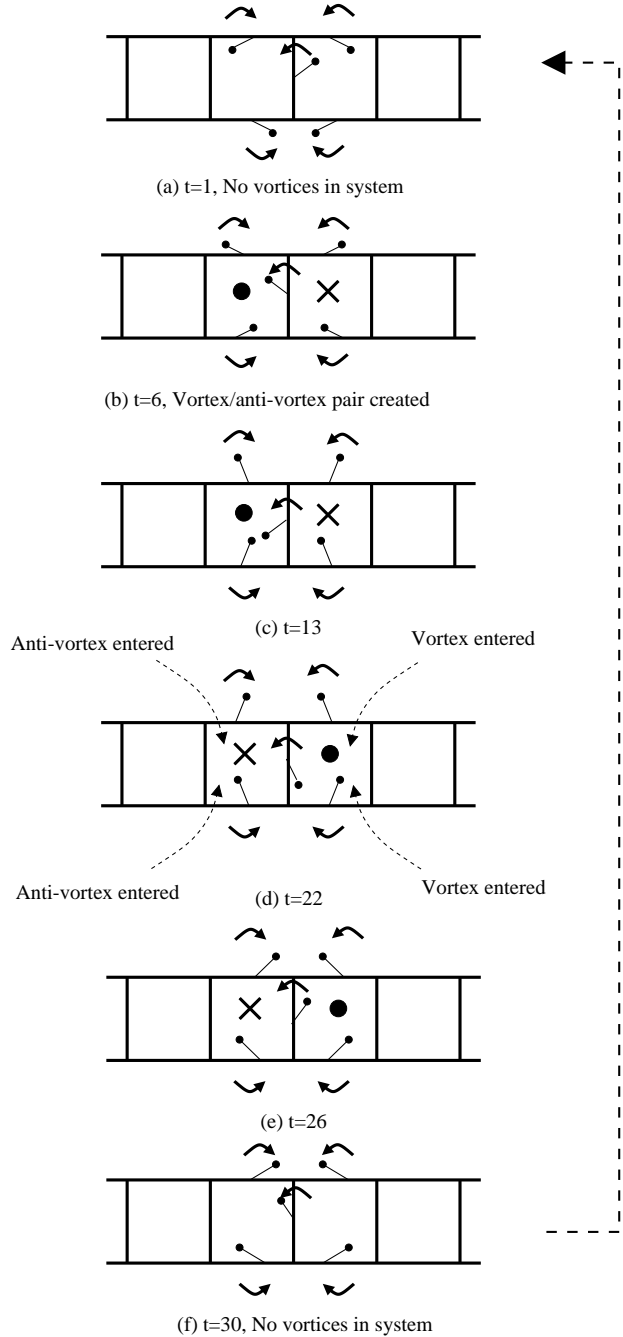}}}
%\vspace{0.15in}
\caption[Vortex pattern of B1$\alpha$ breather for large $\lambda$] {
Vortex pattern of type B1$\alpha$ breather (defined in
fig.~\ref{fig:pltivB}) for large $\lambda$ at $f=0$.}
\label{fig:brmov1B}
\end{figure}

In Fig.~\ref{fig:brmov1B} we show a sequence of time snapshots of the
ladder for a type $B$ breather in the large $\lambda$ limit when
$f=0$.  There is one vertical junction rotating and four horizontal
rotating junctions.  The voltage of the vertical junction is twice of
the horizontal junction.  For every full rotation of a horizontal
junction, the vertical does two full rotations.  The solid circle is a
vortex and the ``$\times$'' is an anti-vortex. In this large
$\lambda$ case, the solution is completely up-down symmetric (see
Fig.~\ref{fig:vt}). Thus $\varphi_j^b(t)=-\varphi_j^t(t)$.

At the initial condition (a) there are no vortices in the ladder.  As
soon as the vertical junction goes over $\pi$ then a vortex-antivortex
pair is created as shown in (b).  This pair remains pinned in the
ladder even though the applied current is large enough to depin
isolated vortices.  When the horizontal junctions go over $\pi$ then
they create two vortices or antivortices as shown in (d).  The result
is as if the vortex and antivortex pair change cells.  This solutions
remains until the vertical junction rotates through $\pi$.  Then,
another vortex-antivortex pair is created that annihilates the pair of
opposite polarity that was in the ladder.  Now, there are no vortices
in the ladder and the sequence repeats itself.  The period double
nature of the solution is evident.  The vertical junction both creates
a vortex-antivortex pair, and also annihilates the vortex-antivortex
pair created by the horizontal junctions.

\begin{figure}[tb]
\epsfxsize=3.2in
\centering{\mbox{\epsfbox{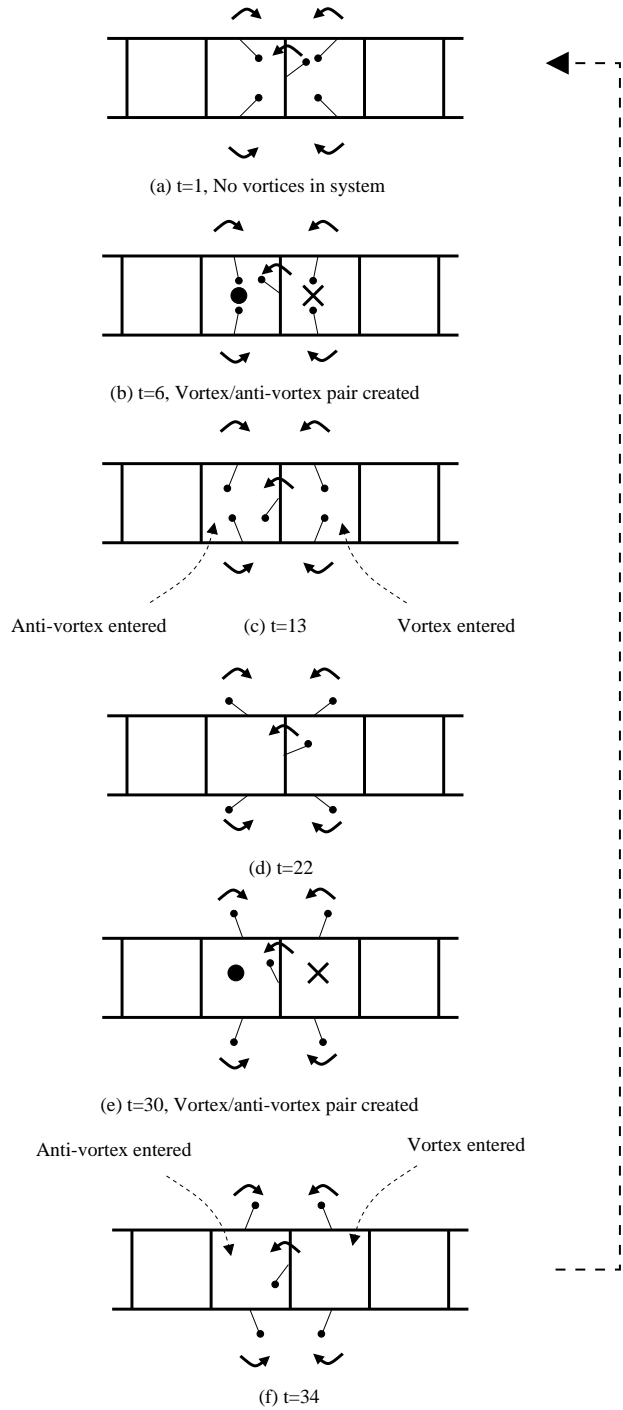}}}
%\vspace{0.15in}
\caption[Vortex pattern of B1 breather for small $\lambda$]
{Vortex pattern of type B1 breather for small $\lambda$
at $f=0$.}
\label{fig:brmov2B}
\end{figure}

\begin{figure}[tb]
\epsfxsize=3in
\centering{\mbox{\epsfbox{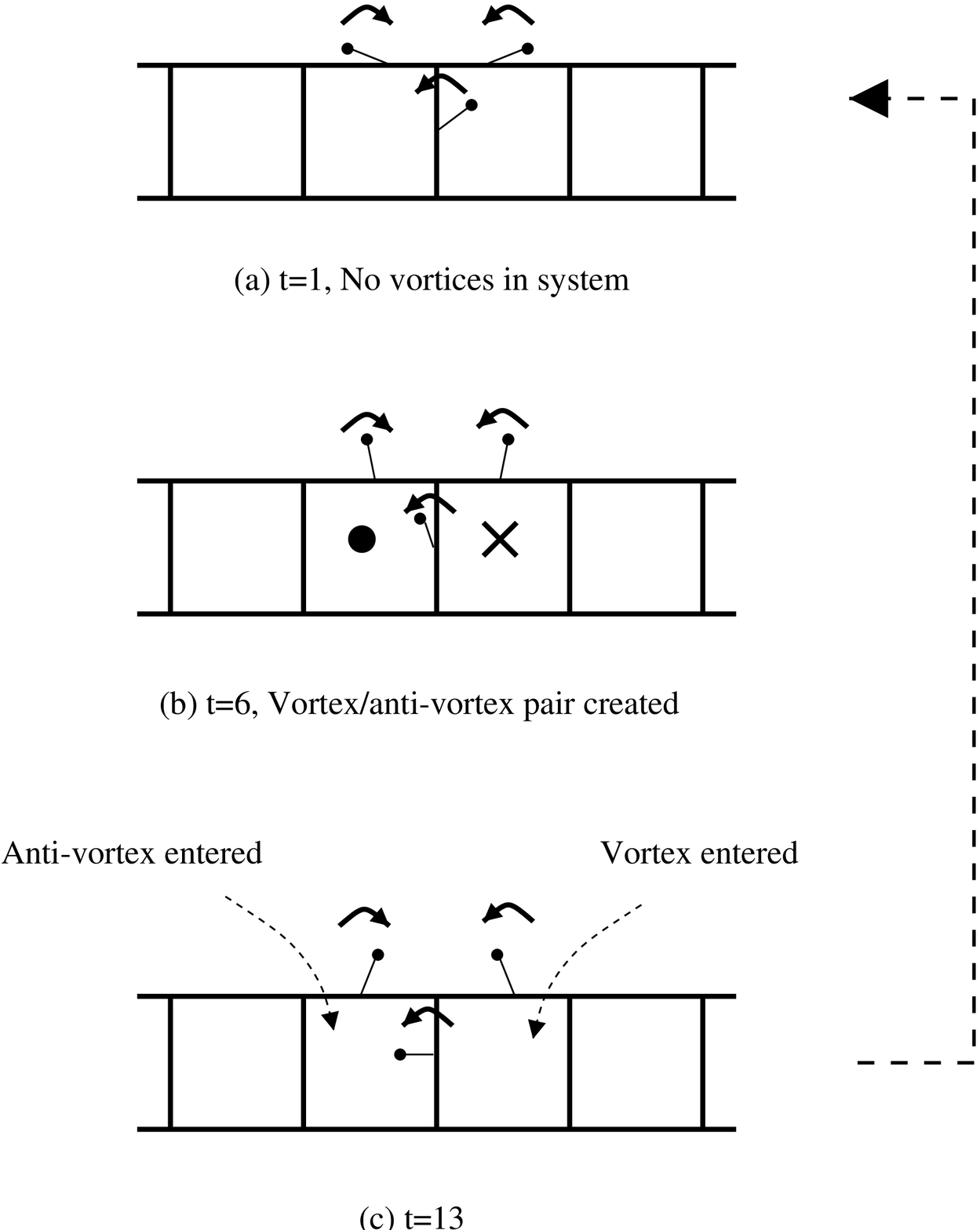}}}
%\vspace{0.15in}
\caption[Vortex pattern of type A breather]
{Vortex pattern of type A breather at $f=0$.}
\label{fig:brmov1A}
\end{figure}

\subsubsection{Small $\lambda$ type B solutions}
\label{sec:vorBsmall}

Fig.~\ref{fig:brmov2B} shows a type B breather in the small $\lambda$
limit at $f=0$.  This solution is similar to Fig.~\ref{fig:brmov1B}
but we no longer have up-down symmetry.  Instead, the top and bottom
rotating horizontal junctions have a $\pi$ phase difference (see
Fig.~\ref{fig:vt}).  We again start when the ladder has no vortices in
(a).  The vertical junction goes over $\pi$ and creates a
vortex-antivortex pair.  Then in (c) the bottom horizontal junctions
go over $\pi$ and annihilate the vortex-antivortex pair.  After some
time, the vertical junction again creates a vortex-antivortex pair
that disappears when the top horizontal junctions go over $\pi$ and
annihilate the vortex-antivortex pair.  

In the small $\lambda$ limit vortex-vortex interactions are 
large.  Therefore, vortices prefer not to enter through the top
and bottom horizontal junctions at the same time.  In order
to minimize this repulsive interaction, vortices and anti-vortices
enter the cells at different times when $\lambda$ is small and
the up-down symmetry of the solution is broken.

\subsubsection{Type A solutions}
\label{sec:vorA}

Fig.~\ref{fig:brmov1A} shows a type A breather for $f=0$.  The
solution in this case is easier to interpret that the type B solutions
since every junction rotates at the same voltage.  First, the vertical
junction goes over $\pi$ and creates a vortex-antivortex.  The
horizontal junctions then rotate over $\pi$ and annihilate the
vortex-antivortex pair as shown in (c).

If we apply a field, then a phase difference develops between the
horizontal junctions.  For instance, in the appropriate field the
right horizontal junction can go over $\pi$ first, thereby injecting a
vortex to the right cell of the rotating vertical junction.  Then, the
vertical junction goes over $\pi$ creating a vortex-antivortex pair.
The antivortex is created in the right cell and the vortex on the
left.  Since there was already a vortex on the right cell, the
anti-vortex annihilates this vortex.  The resulting state is as if the
vortex moved from the right cell to the left cell.  Then, the left
horizontal junction goes over $\pi$ creating an anti-vortex in the
left cell and thereby annihilating the vortex.  This results in the
ladder having no vortices.  The full process can be interpreted as a
rotating vortex around the common node of the rotating junctions.
Therefore, type A breathers in a magnetic field are rotating
vortices.

\section{Discussion and Conclusions}
\label{sec:conbr}

Nonlinearity can cause localization in otherwise perfectly periodic systems.
We have fabricated an experimental system to study these types of localized
solutions called discrete breathers (DB).  We have experimentally detected
different one-site and multi-site DB states in a superconducting Josephson
ladder network.  DB can be excited and annihilated in a controlled way in the
ladder by using a local bias current. Then, by varying the external current
and temperature we have studied the domain of existence and the instability
mechanisms of these localized solutions. Experimentally we find two families
of DB: type B and type A solutions. In the case of type B, four of the
horizontal junctions in the array are in the resistive state, while for type A
only two horizontal junctions are in the resistive state.  We find that some
of the type B and all the type A breathers do not obey the natural up-down
symmetry of the ladder equations. The existence of type A solutions was
predicted in~\cite{mazo99} and recently has also been reported measurements
showing a diversity (type B, type A and hybrid) of DB in an open JJ ladder
array~\cite{binder00b}.

We have developed a series of models and done theoretical analysis and
simulations to explain much of the behavior of the experimental data. At the
simplest level, we have developed a DC circuit model for the system. In this
model the junctions in the resistive state are studied as linear resistors and
the junctions in the superconducting state as shorts. This model allows for
evaluating the effect of the bias circuit and computing IV curves.  We can
estimate the minimum current for which both type A and B breathers can exist
by using this model and assuming that the minimum applied current is
determined by a single junction retrapping mechanism.  We can also predict the
maximum current for which type B breathers can exist by using the junction's
critical current.  All these predictions agree with the experimental results.
However, the type A maximum current found in our experiments is lower than
predicted and is not yet understood.

Using numerical simulations we have found the existence of many different DB
attractors. We find periodic, quasiperiodic and chaotic type B solutions. Such
solutions are identified with different regions in the numerically computed IV
curves, and when we vary the parameters we find many different scenarios for
the behavior of the solutions.  The DB also excite waves in the ladder.  We
have calculated the frequencies of these small waves by using a linear
analysis of the system equations.  A resonance can occur when the DB frequency
matches a frequency of the linear wave.  We have numerically verified that
these resonances can cause instabilities in the localized solutions, but in
most of the cases the resonances are not strong enough for destroying the
localization of the breather.  We have also performed a harmonic balance
approximation which allows the characterization of the amplitudes of type A
and type B breathers.

Since DB do not exist for all the values of the model parameters, we have
studied the regions of existence of localized solutions when these parameters
vary. This study showed good agreement with the predictions of the simple
model in many cases. In brief, DB exist at small values of the anisotropy, but
the underdamped character of the junctions is crucial.  In general, a smaller
$\Gamma$ results in a larger existence region.  Thus, for small values of the
damping DB are predicted to exist even at high values ($>1$) of the anisotropy
parameter.  However, we find a complicated dependence of the DB instabilities
with respect to $\lambda$ which is only partly explained by resonances between
DB frequencies and the lattice linear waves.

When varying $\lambda$ we find different physical behaviors for large and
small values. At large values of $\lambda$, type B solutions are up-down
symmetric, and at small values of $\lambda$ this up-down symmetry is broken.
In the intermediate $\lambda$ regions the resonances between the breather
frequency and the linear modes are important and the behavior is much more
complex. It is in this region where we found chaotic localized solutions.

One important conclusion of these simulations at different values of $\lambda$
is the observation of resonances in the IV curves which do not destroy the DB
solution. This can be understood after the analysis presented in section 6.2.
For the existence of DB, in the case of a forced and damped system the
non-resonance condition is not necessary, because for any frequency the decay
length is different from zero.

The induced magnetic fluxes in the cells of the array librate around non-zero
DC values and have different signs in opposite regions of the array with
respect to the localization of the breather. To complete the physical picture
of the dynamics of the DB we also analyzed the type B and type A solutions in
terms of the vortex-antivortex behavior in the array. Briefly,
vortex-antivortex pairs are created in the center of the array and destroyed
when new vortices and antivortices enter the array through the horizontal
junctions. In the case of a type A solution with an applied external magnetic
field, this vortex picture corresponds to the presence of a rotating vortex.

Many of the aspects of the behavior of the solutions, including ranges
of existence, were understood using simple models.  However
transitions between different states when decreasing the external
current is not fully understood.  Although different behaviors have
been found, a typical experimental result is that type B breathers
become unstable through a cascade of switching events to multi-site
type B solutions. To understand this phenomenon we have done more
detailed simulations of the dynamics of the array. These simulations
are based in the RCSJ model and we have numerically studied the effect
of the bias circuit, the nonlinear resistance, full inductance matrix,
thermal fluctuations and inhomogeneities in the junction
parameters. Although sometimes a switching to a multi-site state was
found, the typical numerical result for small $\lambda$ and $h=0.25$
shows destabilization from type B to type A solutions.  Another aspect
is that all these transitions occurs close to the region of the gap
voltage where it might be necessary to used models more sophisticated
than the RCSJ one.

We have also studied a reduced model where up-down symmetry in the system is
imposed by setting $\varphi_j^t(t)=-\varphi_j^b(t)$.  Then, transitions to
multi-site states are typically found. However this model is unable to explain
the existence of type A solutions and the transitions from type B to type A
solutions that have been reported here and in~\cite{binder00b}. This approach
using a reduced model has been used in~\cite{flach99} and~\cite{giles00} to
study the dynamics of the array.

DB and vortices are two examples of localized excitations in a Josephson
array. Future work will explore the interplay between DB and vortices (kinks
or solitons) solutions which can coexist in the array~\cite{et_thesis}.
Another interesting direction is a experimental study DB solutions in other
geometries such as triangular arrays, and the excitation of breathers by AC
driving currents.

After this article was accepted we learnt of~\cite{giles01} which
reports on numerical work of switching mechanisms in rotobreathers at
$h=0.44$ and $\lambda \simeq 0.21$.

\ack This work was supported by NSF (DMR-9610042 and DMR-9988832),
DGES (PB98-1592), EU (HPRN-CT-1999-00163, LOCNET) and and the
Commission for Cultural, Educational and Scientific Exchange between
the United States of America and Spain. JJM also thanks Terry Orlando
and his group for his hospitality at MIT.

\end{document}

%% file: bias.tex
\expandafter\ifx\csname graph\endcsname\relax \csname newbox\endcsname\graph\fi
\expandafter\ifx\csname graphtemp\endcsname\relax \csname newdimen\endcsname\graphtemp\fi
\setbox\graph=\vtop{\vskip 0pt\hbox{%
    \special{pn 11}%
    \special{pa 108 163}%
    \special{pa 108 325}%
    \special{pa 163 352}%
    \special{pa 54 406}%
    \special{pa 163 460}%
    \special{pa 54 515}%
    \special{pa 163 569}%
    \special{pa 54 623}%
    \special{pa 108 650}%
    \special{pa 108 813}%
    \special{fp}%
    \graphtemp=.5ex\advance\graphtemp by 0.406in
    \rlap{\kern 0.379in\lower\graphtemp\hbox to 0pt{\hss ${ \scriptstyle R_{b}/4}$\hss}}%
    \graphtemp=.5ex\advance\graphtemp by 0.813in
    \rlap{\kern 0.000in\lower\graphtemp\hbox to 0pt{\hss ${ \begin{picture}(10,10) \put(1.75,1.8){\circle{10} } \put(0,0){$\scriptstyle b$}\end{picture} }$\hss}}%
    \special{pa 108 813}%
    \special{pa 271 813}%
    \special{pa 298 758}%
    \special{pa 352 867}%
    \special{pa 406 758}%
    \special{pa 460 867}%
    \special{pa 515 758}%
    \special{pa 569 867}%
    \special{pa 596 813}%
    \special{pa 758 813}%
    \special{fp}%
    \graphtemp=\baselineskip\multiply\graphtemp by -1\divide\graphtemp by 2
    \advance\graphtemp by .5ex\advance\graphtemp by 0.758in
    \rlap{\kern 0.433in\lower\graphtemp\hbox to 0pt{\hss \hbox{$\:$}${ \scriptstyle R_{h} }$\hbox{$\:$}\hss}}%
    \special{pa 758 813}%
    \special{pa 758 650}%
    \special{pa 704 623}%
    \special{pa 813 569}%
    \special{pa 704 515}%
    \special{pa 813 460}%
    \special{pa 704 406}%
    \special{pa 813 352}%
    \special{pa 758 325}%
    \special{pa 758 163}%
    \special{fp}%
    \graphtemp=.5ex\advance\graphtemp by 0.406in
    \rlap{\kern 0.948in\lower\graphtemp\hbox to 0pt{\hss ${ \scriptstyle R_{b} }$\hss}}%
    \graphtemp=.5ex\advance\graphtemp by 0.993in
    \rlap{\kern 0.623in\lower\graphtemp\hbox to 0pt{\hss ${ \begin{picture}(10,10) \put(1.75,1.8){\circle{10} } \put(0,0){$\scriptstyle c$}\end{picture} }$\hss}}%
    \special{pa 758 163}%
    \special{pa 108 163}%
    \special{fp}%
    \graphtemp=.5ex\advance\graphtemp by 0.000in
    \rlap{\kern 0.840in\lower\graphtemp\hbox to 0pt{\hss ${ \begin{picture}(10,10) \put(1.75,1.6){\circle{10}} \put(0,0){$\scriptstyle a$}\end{picture} }$\hss}}%
    \special{pa 758 163}%
    \special{pa 1408 163}%
    \special{fp}%
    \special{pa 1408 163}%
    \special{pa 1408 325}%
    \special{pa 1463 352}%
    \special{pa 1354 406}%
    \special{pa 1463 460}%
    \special{pa 1354 515}%
    \special{pa 1463 569}%
    \special{pa 1354 623}%
    \special{pa 1408 650}%
    \special{pa 1408 813}%
    \special{fp}%
    \graphtemp=.5ex\advance\graphtemp by 0.406in
    \rlap{\kern 1.679in\lower\graphtemp\hbox to 0pt{\hss ${ \scriptstyle R_{b}/4}$\hss}}%
    \special{pa 1408 813}%
    \special{pa 1246 813}%
    \special{pa 1219 867}%
    \special{pa 1165 758}%
    \special{pa 1110 867}%
    \special{pa 1056 758}%
    \special{pa 1002 867}%
    \special{pa 948 758}%
    \special{pa 921 813}%
    \special{pa 758 813}%
    \special{fp}%
    \graphtemp=\baselineskip\multiply\graphtemp by -1\divide\graphtemp by 2
    \advance\graphtemp by .5ex\advance\graphtemp by 0.758in
    \rlap{\kern 1.083in\lower\graphtemp\hbox to 0pt{\hss \hbox{$\:$}${ \scriptstyle R_{h} }$\hbox{$\:$}\hss}}%
    \special{pa 1408 813}%
    \special{pa 1408 1463}%
    \special{fp}%
    \special{pa 1408 1463}%
    \special{pa 1246 1463}%
    \special{pa 1219 1517}%
    \special{pa 1165 1408}%
    \special{pa 1110 1517}%
    \special{pa 1056 1408}%
    \special{pa 1002 1517}%
    \special{pa 948 1408}%
    \special{pa 921 1463}%
    \special{pa 758 1463}%
    \special{fp}%
    \graphtemp=\baselineskip\multiply\graphtemp by -1\divide\graphtemp by 2
    \advance\graphtemp by .5ex\advance\graphtemp by 1.408in
    \rlap{\kern 1.083in\lower\graphtemp\hbox to 0pt{\hss \hbox{$\:$}${ \scriptstyle R_{h}}$\hbox{$\:$}\hss}}%
    \special{pa 758 1463}%
    \special{pa 758 1300}%
    \special{pa 704 1273}%
    \special{pa 813 1219}%
    \special{pa 704 1165}%
    \special{pa 813 1110}%
    \special{pa 704 1056}%
    \special{pa 813 1002}%
    \special{pa 758 975}%
    \special{pa 758 813}%
    \special{fp}%
    \graphtemp=.5ex\advance\graphtemp by 1.056in
    \rlap{\kern 0.948in\lower\graphtemp\hbox to 0pt{\hss ${ \scriptstyle R_{v}}$\hss}}%
    \special{pa 108 813}%
    \special{pa 108 1463}%
    \special{fp}%
    \special{pa 108 1463}%
    \special{pa 271 1463}%
    \special{pa 298 1408}%
    \special{pa 352 1517}%
    \special{pa 406 1408}%
    \special{pa 460 1517}%
    \special{pa 515 1408}%
    \special{pa 569 1517}%
    \special{pa 596 1463}%
    \special{pa 758 1463}%
    \special{fp}%
    \graphtemp=\baselineskip\multiply\graphtemp by -1\divide\graphtemp by 2
    \advance\graphtemp by .5ex\advance\graphtemp by 1.408in
    \rlap{\kern 0.433in\lower\graphtemp\hbox to 0pt{\hss \hbox{$\:$}${ \scriptstyle R_{h} }$\hbox{$\:$}\hss}}%
    \special{pa 108 1463}%
    \special{pa 108 1625}%
    \special{pa 163 1652}%
    \special{pa 54 1706}%
    \special{pa 163 1760}%
    \special{pa 54 1815}%
    \special{pa 163 1869}%
    \special{pa 54 1923}%
    \special{pa 108 1950}%
    \special{pa 108 2113}%
    \special{fp}%
    \graphtemp=.5ex\advance\graphtemp by 1.788in
    \rlap{\kern 0.163in\lower\graphtemp\hbox to 0pt{\hbox{$\:$}${ \scriptstyle R_{b}/4}$\hbox{$\:$}\hss}}%
    \special{pa 108 2113}%
    \special{pa 758 2113}%
    \special{fp}%
    \special{pa 758 2113}%
    \special{pa 758 1950}%
    \special{pa 704 1923}%
    \special{pa 813 1869}%
    \special{pa 704 1815}%
    \special{pa 813 1760}%
    \special{pa 704 1706}%
    \special{pa 813 1652}%
    \special{pa 758 1625}%
    \special{pa 758 1463}%
    \special{fp}%
    \graphtemp=.5ex\advance\graphtemp by 1.788in
    \rlap{\kern 0.813in\lower\graphtemp\hbox to 0pt{\hbox{$\:$}${ \scriptstyle R_{b}}$\hbox{$\:$}\hss}}%
    \special{pa 1408 1463}%
    \special{pa 1408 1625}%
    \special{pa 1463 1652}%
    \special{pa 1354 1706}%
    \special{pa 1463 1760}%
    \special{pa 1354 1815}%
    \special{pa 1463 1869}%
    \special{pa 1354 1923}%
    \special{pa 1408 1950}%
    \special{pa 1408 2113}%
    \special{fp}%
    \graphtemp=.5ex\advance\graphtemp by 1.788in
    \rlap{\kern 1.463in\lower\graphtemp\hbox to 0pt{\hbox{$\:$}${\scriptstyle R_{b}/4}$\hbox{$\:$}\hss}}%
    \special{pa 1408 2113}%
    \special{pa 758 2113}%
    \special{fp}%
    \hbox{\vrule depth2.113in width0pt height 0pt}%
    \kern 1.679in
  }%
}%

%% file: trias.bbl
\begin{thebibliography}{10}

\bibitem{trias00} E.~Tr\'{\i}as, J.~J. Mazo and T.~P. Orlando.
Discrete breathers in nonlinear lattices: Experimental detection in a
{Josephson} array, {\em Phys.\ Rev.\ Lett.} {\bf 84} (2000) 741--744.

\bibitem{binder00a} P.~Binder, D.~Abraimov, A.~V. Ustinov, S.~Flach,
and Y.~Zolotaryuk. Observation of breathers in Josephson ladders,
{\em Phys.\ Rev.\ Lett.} {\bf 84} (2000) 745--748.

\bibitem{binder00b} P.~Binder, D.~Abraimov and A.~V. Ustinov.
Diversity of discrete breathers observed in a Josephson
ladder, {\em Phys.\ Rev.\ E} {\bf 62} (2000) 2858--2862.

\bibitem{sievers88} A.~J.~Sievers and S.~Takeno. Intrinsic localized
modes in anharmonic crystals, {\em Phys.\ Rev.\ Lett.} {\bf 61}
(1988) 970--973.

\bibitem{mackay94} R.~S. MacKay and S.~Aubry. Proof of existence of
breathers for time-reversible or hamiltonian networks of weakly
coupled oscillators, {\em Nonlinearity} {\bf 7} (1994) 1623--1643.

\bibitem{takeno96} S.~Takeno and M.~Peyrard. Nonlinear modes in
coupled rotator models, {\em Physica D} {\bf 55} (1996) 140--163.

\bibitem{mackay98} R.~S. MacKay and J.-A. Sepulchre. Stability of
discrete breathers, {\em Physica D} {\bf 119} (1998) 148--162.

\bibitem{martinez99} P.~J. Mart\'{\i}nez, L.~M. Flor\'{\i}a, F. Falo
and J.~J. Mazo. Intrinsically localized chaos in discrete nonlinear
extended systems, {\em Europhys.\ Lett.} {\bf 45} (1999) 444--449.

\bibitem{tsironis96} G.~P. Tsironis and S. Aubry. Slow relaxation
phenomena induced by breathers in nonlinear lattices, {\em Phys.\
Rev.\ Lett.} {\bf 77} (1996) 5225--5228.

\bibitem{archilla99} J.~F.~R. Archilla, R.~S. MacKay and
J.~L. Mar\'{\i}n. Discrete breathers and Anderson modes: two faces of
the same phenomenon?, {\em Physica D} {\bf 134} (1999) 406--418.

\bibitem{kopidakis00} G. Kopidakis and S. Aubry. Discrete breathers
and delocalization in nonlinear disordered systems, {\em Phys.\ Rev.\
Lett.} {\bf 84} (2000) 3236--3239.

\bibitem{aubry97} S. Aubry. Breathers in nonlinear lattices:
Existence, linear stability and quantization, {\em Physica D} {\bf 103}
(1997) 201--250.

\bibitem{flach98} S. Flach and  C.~R. Willis. Discrete breathers,
{\em Phys.\ Rep.} {\bf 295} (1998) 182-264.

\bibitem{sievers95} A.~J.~Sievers and J.~B.~Page. Unusual anharmonic local
mode systems, in: G.~K.~Horton and A.~A.~Maradudin, eds. {\em Dynamical
Properties of Solids VII} (Elsevier, Amsterdam, 1995) 137--

\bibitem{floria96} L.~M. Flor\'{\i}a, J.~L. Mar\'{\i}n,
P.~J. Mart\'{\i}nez, F.~Falo and S.~Aubry. Intrinsic localisation in
the dynamics of a Josephson-junction ladder, {\em Europhys.\ Lett.}
{\bf 36} (1996) 539--544.

\bibitem{mcgurn99} A.~R.~McGurn. Intrinsic localized modes in
nonlinear photonic crystal waveguides: Dispersive modes, {\em Phys.\
Letts.\ A} {\bf 260} (1999) 314--321.

\bibitem{lai99} R.~Lai and A.~J.~Sievers. Nonlinear nanoscale
localization of magnetic excitations in atomic lattices. {\em Phys.\ Rep.}
{\bf 314} (1999) 147--236.

\bibitem{vakakis96} J.~Aubrecht, A.~F. Vakakis, T.-C. Tsao and
J.~Bentsman. Experimental study of non-linear transient motion
confinement in a system of coupled beams, {\em J.\ Sound Vib.}
{\bf 195} (1996) 629--648.

\bibitem{vakakis99} A.~F. Vakakis, A.~N. Kounadis and
I.~G. Raftoyiannis. Use of non-linear localization for isolating
structures from earthquake-induced motions, {\em Earthquake Eng.\
Struc.} {\bf 28} (1999) 21--36.

\bibitem{levine93} M.~B. Levine-West and M.~A. Salama.  Mode
localization experiments on a ribbed antenna, in: {\em Proc. 33rd
AIAA/ASME Structures Dynamics and Materials Conference}
{\bf 31} (1993) 1929--1937.

\bibitem{peyrard98} M.~Peyrard.
Using DNA to probe nonlinear localized excitations?,
{\em Europhys.\ Lett.} {\bf 44} (1998) 271--277.

\bibitem{swanson99} B.~I.~Swanson, J.~A.~Brozik, S.~P.~Love,
G.~F.~Strouse, A.~P.~Shreve, A.~R.~Bishop, W.-Z.~Wang and
M.~I.~Salkola. Observation of Intrinsically Localized Modes in a
Discrete Low-Dimensional Material.  {\em Phys.\ Rev.\ Lett.} {\bf 82}
(1999) 3288-3291.

\bibitem{schwarz99} U.~T. Schwarz, L.~Q. English and A.~J. Sievers.
Experimental Generation and Observation of Intrinsic Localized Spin
Wave Modes in an Antiferromagnet, {\em Phys.\ Rev.\ Lett.} {\bf 83}
(1999) 223-226.

\bibitem{floria98} L.~M. Flor\'{\i}a, J.~L. Mar\'{\i}n, S. Aubry,
P.~J. Mart\'{\i}nez, F. Falo and J.~J. Mazo. Josephson-junction
ladder: A benchmark for nonlinear concepts, {\em Physica D} {\bf 113}
(1998) 387--396.

\bibitem{martinez98} P.~J. Mart\'{\i}nez, L.~M. Flor\'{\i}a,
J.~L. Mar\'{\i}n, S. Aubry and J.~J. Mazo. Floquet stability of
discrete breathers in anisotropic Josephson junction ladders, {\em
Physica D} {\bf 119} (1998) 175--183.

\bibitem{flach99} S.~Flach and M.~Spicci. Rotobreather dynamics in
underdamped josephson junction ladders, {\em J.\ Phys. Cond.\ Matter}
{\bf 11} (1999) 321--334.

\bibitem{mazo99} J.~J. Mazo, E.~Tr\'{\i}as and T.~P. Orlando.
Discrete breathers in dc-biased Josephson-junction arrays, {\em Phys.\
Rev.\ B} {\bf 59} (1999) 13604--13607.

\bibitem{likharev_book} K.~K. Likharev, {\em Dynamics of Josephson
junctions and circuits}, Gordon and Breach Science Publishers,
Pennsylvania, 1986.

\bibitem{floria96fk} L.~M.~Flor\'{\i}a and J.~J~Mazo. Dissipative
dynamics of the Frenkel-Kontorova model, {\em Adv.\ Phys.} {\bf 45} (1996)
505--598.

\bibitem{XY} D.~J. Resnick, J.~C. Garland, J.~T. Boyd, S. Shoemaker
and R.~S. Newrock. Kosterlitz-Thouless transition in proximity-coupled
superconducting array, Phys.\ Rev.\ Lett. {\bf 47} (1981) 1542--1545.

\bibitem{watanabe95}
Shinya Watanabe, S.~H. Strogatz, H.~S.~J. van~der Zant and T.~P. Orlando.
Whirling modes and parametric instabilities in the discrete
sine-Gordon equation: Experimental tests in Josephson rings,
{\em Phys.\ Rev.\ Lett.}, {\bf 74} (1995) 379--382.

\bibitem{mazo95} J.~J. Mazo, F. Falo and Luis~M. Flor\'{\i}a.
Josephson-junction ladders: Ground state and relaxation phenomena,
{\em Phys.\ Rev.\ B} {\bf 52} (1995) 10433--1040.

\bibitem{unpublished} J.~J.~Mazo, unpublished.

\bibitem{Hypres} {Hypres Inc.} Elmsfort, {NY} 10523, superconducting
circuits foundry.

\bibitem{henry} FastHenry, ftp://rle-vlsi.mit.edu/pub/fasthenry.

\bibitem{chen88} Y.~C. Chen, M.~P.~A. Fisher and A.~J. Leggett. The
return of a hysteretic Josephson junction to the zero-voltage
state--IV characteristic and quantum retrapping.  {\em J.\ Appl.\
Phys.} {\bf 64} (1988) 3119--3142.

\bibitem{kirtley88} J.~R. Kirtley, C.~D. Tesche, W.~J. Gallagher,
A.~W. Kleinsasser, R.~L.  Sandstrom, S.~I. Raider and
M.~P.~A. Fisher. Measurement of the intrinsic subgap dissipation in
Josephson junctions, {\em Phys.\ Rev.\ Lett.}  {\bf 61} (1988)
2372--2375.

\bibitem{cristiano97} R.~Cristiano, L.~Frunzio, C.~Nappi,
M.~G. Castellano, G.~Torrioli and C.~Cosmelli.  The effective
dissipation in Nb/AlO$_x$/Nb {Josephson} tunnel junctions by return
current measurements, {\em J.\ Appl.\ Phys.} {\bf 81} (1997) 7418--7426.

\bibitem{castellano99} M.~G. Castellano, G.~Torrioli, F.~Chiarello,
C.~Cosmelli and P.~Carelli. Return current in hysteretic Josephson
junctions: Experimental distribution in the thermal activation regime,
{\em J.\ Appl.\ Phys.} {\bf 86} (1999) 6405--6411.

\bibitem{giles00} R.~T.~Giles and F.~V.~Kusmartsev. Switching and
symmetry breaking behaviour of discrete breathers in Josephson
ladders, {\em cond-mat/0004078}.

\bibitem{et_thesis} E. Tr\'{\i}as. Ph. D. Thesis, MIT, 2000.

\bibitem{giles01} R.~T.~Giles and F.~V.~Kusmartsev. Chaotic transients
in the switching of roto-breathers, {\em cond-mat/0101439}.

\end{thebibliography}
